\documentclass[oldversion,preprint]{aa}  
\usepackage{graphicx}
\usepackage{txfonts}
\usepackage{longtable}
\usepackage{multirow}

\begin{document}

\title{Detection of jet precession in the active nucleus of M81}

\titlerunning{Detection of jet precession in M81*}

   \author{I. Mart\'i-Vidal\inst{1}
          \and
          J. M. Marcaide\inst{2}
          \and
          A. Alberdi\inst{3}
          \and
          M. A. P\'erez-Torres\inst{3}
          \and
          E. Ros\inst{2,1}
          \and
          J. C. Guirado\inst{2}
}

   \institute{
             Max-Planck-Institut f\"ur Radioastronomie,
             Auf dem H\"ugel 69, D-53121 Bonn (Germany) 
             \email{imartiv@mpifr-bonn.mpg.de}
         \and
            Dpt. Astronomia i Astrof\'isica, 
            Universitat de Val\`encia,
            C/ Dr. Moliner 50, E-46100 Burjassot (Spain)
         \and
            Instituto de Astrof\'isica de Andaluc\'ia (CSIC)
            Apdo. Correos 2004, 08071 Granada (Spain)
}

   \date{Accepted for publication in A\&A (in press).}
 
  \abstract
{
We report on very-long-baseline-interferometry (VLBI) monitoring observations of the low-luminosity active 
galactic nucleus (LLAGN) in the galaxy M\,81 at the frequencies of 1.7, 2.3, 5.0, and 8.4\,GHz. The observations 
reported here are phase-referenced to the supernova SN\,1993J (located in the same galaxy) and 
cover from late 1993 to late 2005. The large amount of available observations allows us to study the 
stability of the AGN position in the
frame of its host galaxy at different frequencies and chromatic effects in the jet morphology, 
together with their time evolution. The source consists at all frequencies of a slightly resolved core and a 
small jet extension towards the northeast direction (position angle of $\sim$65 degrees) in agreement 
with previous publications. We find that the position of the intensity peak in the images 
at 8.4\,GHz is very stable in the galactic frame of M\,81 (proper motion upper limit about
10$\mu$as per year). We confirm previous reports that the peaks at all frequencies are systematically 
shifted among them, possibly due to opacity effects in the jet as predicted by the standard relativistic 
jet model. We use this model, under plausible assumptions, to estimate the magnetic field in 
the jet close to the jet base and the mass of the central black hole. We obtain a black-hole mass of 
$\sim$$2\times10^7$\,M$_{\odot}$, comparable to estimates previously reported using different approaches, 
but the magnetic fields obtained are $10^3-10^4$ times lower than previous estimates. We find that the 
positions of the cores at 1.7, 2.3, and 
5.0\,GHz are less stable than that at 8.4\,GHz and evolve systematically, shifting southward at a rate of 
several tens of $\mu$as per year. The evolution in the jet orientation seems to be related to changes in 
the inclination of the cores at all frequencies. These results can be 
interpreted as due to a precessing jet. The evolving jet orientation also seems to be related 
to a flare in the peak flux 
densities at 5.0 and 8.4\,GHz, which lasts $\sim$4 years (from mid 1997 to mid 2001). An increase in the 
accretion rate of the black hole, and its correlation with the jet luminosity via the disk-jet connection
model, seems insufficient to explain this long flare and
the simultaneous evolution in the jet orientation. A continued monitoring of the flux density and the jet 
structure evolution in this LLAGN will be necessary to further confirm our jet precession model.}
   \maketitle

\section{Introduction}
\label{I}

The galaxy M\,81 (NGC\,3031), located at a distance of $3.63\pm0.34$\,Mpc
(Freedman et al. \cite{Freedman1994}), hosts the next closest active galactic
nucleus (AGN) after Centaurus A. Its high declination (J2000.0 coordinates
$\alpha = 09^\mathrm{h}\,55^\mathrm{m}\,33.173^\mathrm{s}$ and
$\delta = 69^{\circ}\,03'\,55.062''$) also makes M\,81 a perfect target
for global VLBI observations using the most sensitive radio 
telescopes in the Northern Hemisphere.

The flux density of the AGN in M\,81 (hereafter, M\,81*), has
relatively large variability, from X-rays (Ishisaki et al. \cite{Ishisaki1996};
Page et al. \cite{Page2004}) to radio wavelengths (Ho et al. \cite{Ho1999}
and Bietenholz, Bartel \& Rupen \cite{Bietenholz2000}, at centimeter
wavelengths; Sch\"odel et al. \cite{Schodel2008}, at millimeter wavelengths).
The variability time scale runs from years to days, indicating a contribution from
a very compact source to the emission at all frequencies. Indeed, even from VLBI
observations at 43\,GHz (Ros \& P\'erez-Torres \cite{Ros2008}) and space-VLBI
observations at 5\,GHz (Bartel \& Bietenholz \cite{Bartel2000}) the emitting
structure could not be fully resolved.
The broadband luminosity of M\,81* is relatively weak (see, e.g., Ho et al. \cite{Ho1996})
from radio (around 100\,mJy, i.e., on the order of $10^{37}$\,erg\,s$^{-1}$)
to the optical and X-rays ($\sim10^{40}$\,erg\,s$^{-1}$). Therefore, it is
classified as a low-luminosity AGN (LLAGN).

The accretion mechanisms in LLAGNs are not clear. Since relativistic
X-ray lines are not clearly detected in these objects (e.g., Reynolds et al. 
\cite{Reynolds2009}), radial flows or advection mechanisms might dominate the 
accretion, in contrast to the more luminuos AGN. In any case, if the X-ray
emission traces the innermost inflow from the disk to the central engine and 
the radio emission traces the jet power, there should be a relationship between 
luminosities in both X-rays and radio for all AGNs (i.e., including LLAGNs), which 
should also be related to the accretion rate. This is the basis of the 
{\em fundamental-plane} model of black-hole accretion (e.g., Merloni, Heinz, \& di 
Matteo \cite{Merloni2003}). 
However, the parametric space of very-low accretion flows in AGNs (e.g.,
LLAGN) is not fully characterized observationally. Although 
Markoff et al. (\cite{Markoff2008}) and Miller et al. (\cite{Miller2010}) were 
able to fit average broadband spectra of M\,81* in the frame of the fundamental-plane 
model of disk-jet connection, no clear correlation between the variabilities in 
the X-rays and radio was observed unlike in microquasars 
(Mirabel et al. \cite{Mirabel1998}) and AGNs (Marscher et al. \cite{Marscher2002}). 
In the present paper, we report on a very long 
flare of M\,81* at radio frequencies, beginning around mid 1997, which lasted
nearly four years. As shown in this paper, an intriguing result 
related to this flare is that it seems to be related to the evolution in the jet 
orientation, thus making it difficult explain in the frame of the disk-jet 
connection between accretion flow and jet power.

At radio frequencies, M\,81* has a slightly inverted spectrum with an spectral index 
$\alpha = +0.3$ ($S\propto\nu^{+\alpha}$), up to a turnover frequency of $\sim$200\,GHz 
(Reuter \& Lesch \cite{Reuter1996}). 
If due to synchrotron self-absorption in the jet, such a high turnover 
frequency, suggests a very dense plasma and a high magnetic field, 
and/or a rather flat energy distribution
of the relativistic synchrotron-emitting particles (see, e.g., Sect. 4.1 of Reuter 
\& Lesch \cite{Reuter1996}). {In addition, it is worth noticing that} 
the spectral
behavior of M\,81* is similar to that of the central engine in our Galaxy,
Sgr\,A* (Duschl \& Lesch \cite{Duschl1994}), thus suggesting that similar
physical mechanisms should be present in the acceleration of the relativistic 
particles and their emission at radio frequencies. In addition, 
the mass of its associated black hole ($7^{+2}_{-1}\times10^{7}$\,M$_{\odot}$, 
Devereux et al. \cite{Devereux2003}; $5.5^{+3.6}_{2.0}\times10^{7}$\,M$_{\odot}$, Schorr M\"uller 
et al. \cite{Schorr2011}) makes M\,81* an interesting intermediate 
object between Sgr A* and the more powerful AGN.

The explosion of the radio-luminous supernova SN\,1993J in M\,81,
which happend around 28 March 1993 (e.g., Weiler et al. \cite{Weiler2007}),
triggered an intense campaign of VLBI observations in which M\,81* was selected as 
the phase calibrator of the SN\,1993J visibilities.
These observations allow us to study the evolution in the structure and position
of M\,81*, in the frame of its host galaxy and at different frequencies and
times.
Results for some of the M\,81* observations at 8.4\,GHz between years 1993 and 1997 are
reported in Bietenholz, Bartel \& Rupen (\cite{Bietenholz2000}), and results
for the astrometry analysis of a subset of observations at all frequencies between years 1993
and 2002 were reported in Bietenholz
et al. (\cite{Bietenholz2004}). These authors conclude that the proper motion of M\,81*,
relative to the shell center of SN\,1993J, was consistent with zero up to the precision
level of the observations ($11.4 \pm 9.3$\,$\mu$as\,yr$^{-1}$). The contributions of galactic
rotation and/or any peculiar motion of the SN\,1993J progenitor were also discarded to
contribute to a detectable proper motion of M\,81* relative to the supernova.
These authors also report a frequency-dependent shift of the peak of brightness of M\,81*,
which they interpret as opacity effects in the jet of the AGN, in agreement with
the standard relativistic jet model (Blandford \& K\"onigl \cite{Blandford}).

In this paper, we report on the analysis of all the VLBI
observations of M\,81* taken in the phase-referencing observing campaign of
SN\,1993J, from late 1993 to late 2005.
In Sect. \ref{ObsSecVLBI} we describe our observations. In Sects. \ref{StrucSec} and
\ref{ResultsSecII}
we report the main results obtained (source morphology and multi-frequency astrometry). 
In Sect. \ref{DiscussionSec}, we analyze the astrometry results in terms of the standard 
jet interaction model (Blandford \& K\"onigl \cite{Blandford}), and in the last section we discuss the detection 
of a changing jet orientation, which we interpret as due to jet precession.

\section{Observations and data reduction}
\label{ObsSecVLBI}

We analyzed the VLBI data from the observing campaigns led by N. Bartel 
(York University, Canada) and J.M. Marcaide (University of Valencia, Spain), from late 1993 to late
2005 (see Table 1 of Mart\'i-Vidal et al. \cite{PaperI} and Table 1 of this paper). SN\,1993J 
was observed using 
M\,81* as the phase calibrator at the frequencies of 1.7, 2.3, 5.0, 8.4, 15, and 
22\,GHz. (There are only a few low-quality observations during years 1993 and 1994 
at the two highest frequencies, which have not been used in our analysis.) The 
results of this observing campaign of SN\,1993J 
are reported in several publications (e.g., Bartel et al. \cite{Bartel2002}; 
Marcaide et al. \cite{Marcaide2009}; Mart\'i-Vidal et al. \cite{PaperI},\cite{PaperII}; 
and references therein), 
where the technical details of all these observations are given. Basically, 
one epoch typically lasted $\sim$12\,h, with the participation 
of 10--15 antennas. The whole Very Long Baseline Array (VLBA), the Very Large 
Array (VLA) in phased-array mode, and some stations of the European VLBI Network 
(EVN) took part in most of the observations. The recording 
rate was set to either 128 or 256 Mbit\,s$^{-1}$, depending on the technical 
capabilities of the recording systems at each epoch. 

\addtocounter{table}{1}

To obtain the calibrated visibilities of M\,81* at each epoch, we proceeded as described in 
Mart\'i-Vidal et al. (\cite{PaperI}). The Astronomical Image Processing System 
({\sc aips}) of the National Radio Astronomy Observatory (NRAO) was used to 
calibrate the data.
Fringe-finder scans were used to align the phases between the different 
observing sub-bands and, afterwards, global fringe fitting (Alef \& Porcas \cite{Alef1986}) 
was applied to the M\,81* visibilities.  The resulting gains
were interpolated to the SN\,1993J scans. The amplitude calibration
was based on gain curves and system temperatures measured at each station, 
with the exception of the phased VLA (where estimates of the source flux densities
are necessary, since the amplitude calibration is based on measurements 
of the ratio of effective antenna temperature to system temperature). Then, 
a refinement of the amplitude gain estimates was performed via self-calibration 
of the M\,81* visibilities, using the {\sc aips} task CALIB.

\section{Results for source morphology}
\label{StrucSec}

\begin{figure*}[ht!]
\centering
\includegraphics[width=18cm]{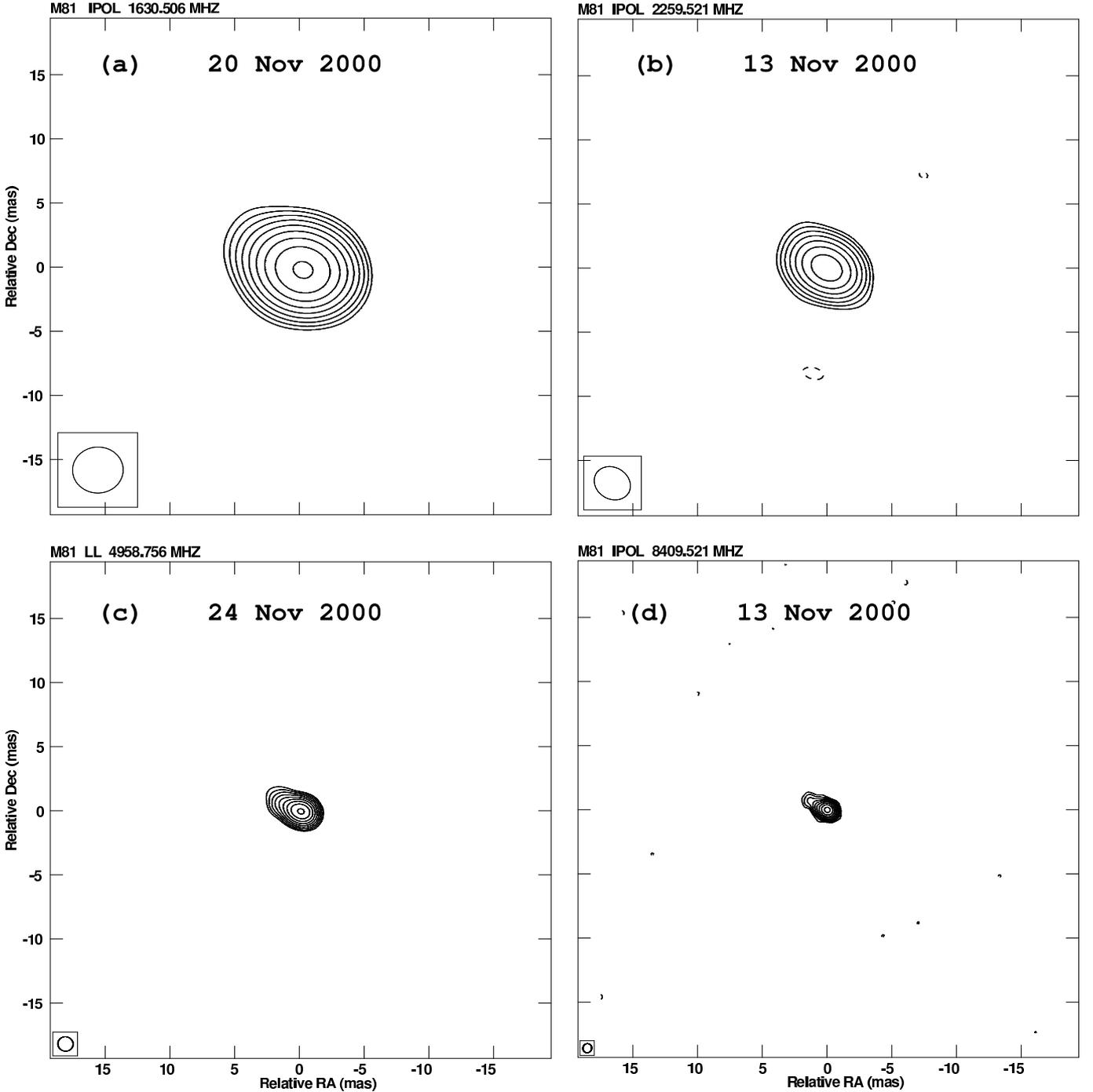}
\caption{Images of M\,81* obtained from the VLBI epochs at (a) 1.6\,GHz, (b) 2.3\,GHz, (c) 5.0\,GHz, 
and (d) 8.4\,GHz observed around mid November 2000 (see Table 1). Contours are located at 
3, $3\sqrt{3}$, 9, ... of the rms of the images. The 
FWHM of the convolving beam is shown in the bottom left corner of each image.}
\label{VLBIMaps}
\end{figure*}

We show in Fig. \ref{VLBIMaps} some representative 
examples of the VLBI images obtained at all the observing frequencies. 
The images of M\,81* consist of a compact source, which dominates the emission, 
and a weak extension towards the northeast. These images agree with 
the results reported in Bietenholz, Bartel, \& Rupen (\cite{Bietenholz2000}).

\subsection{Model fitting}
\label{ModFitSec}

For the analysis of evolution in the M\,81* structure, we characterized the model emission at each 
epoch in a similar way to what was modeled in Bietenholz, Bartel, \& Rupen (\cite{Bietenholz2000}). 
We used an elliptical Gaussian to model the central dominant emission and a point source to model 
the weak jet extension. For those epochs in which the point-source model could not describe the 
M\,81* jet extension satisfactorily, we used a circularly symmetric Gaussian. The point source 
used to succesfully model the jet extension at some epochs is indicative of a compact emission away 
from the core (i.e.,
a hot spot in the jet), at least up to the sensitivity limit achieved in the images. 

\begin{figure*}[ht!]
\centering
\includegraphics[width=18cm]{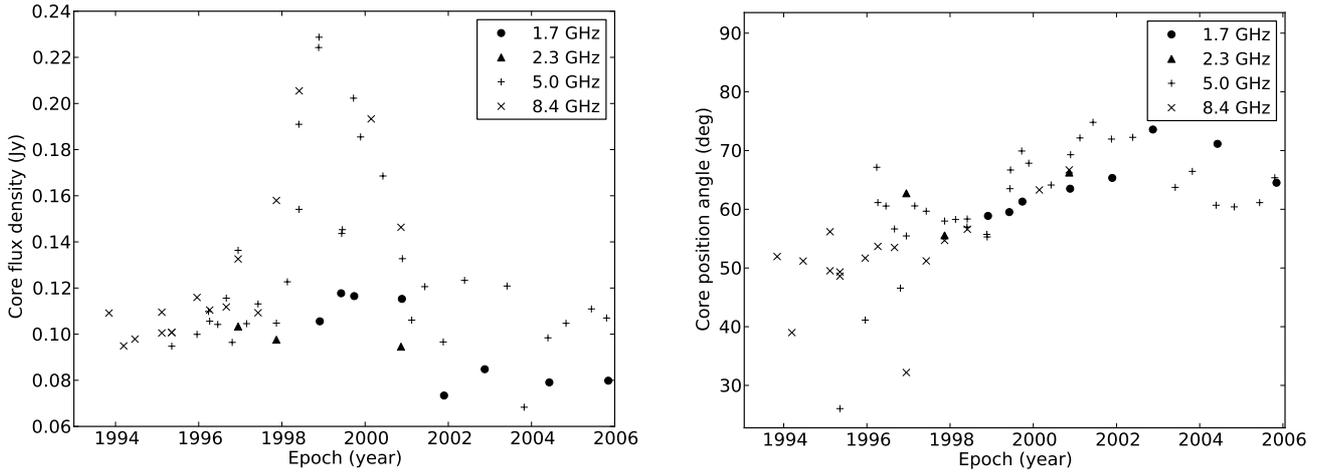}
\caption{Peak flux density (left) and position angle (right) of the elliptical Gaussians used to model 
the core emission of M\,81* at all frequencies (see Table 1 and text). Typical uncertainties are given
in Table \ref{UncertTable}.}
\label{CoreMod1}
\end{figure*}

\begin{figure*}[ht!]
\centering
\includegraphics[width=18cm]{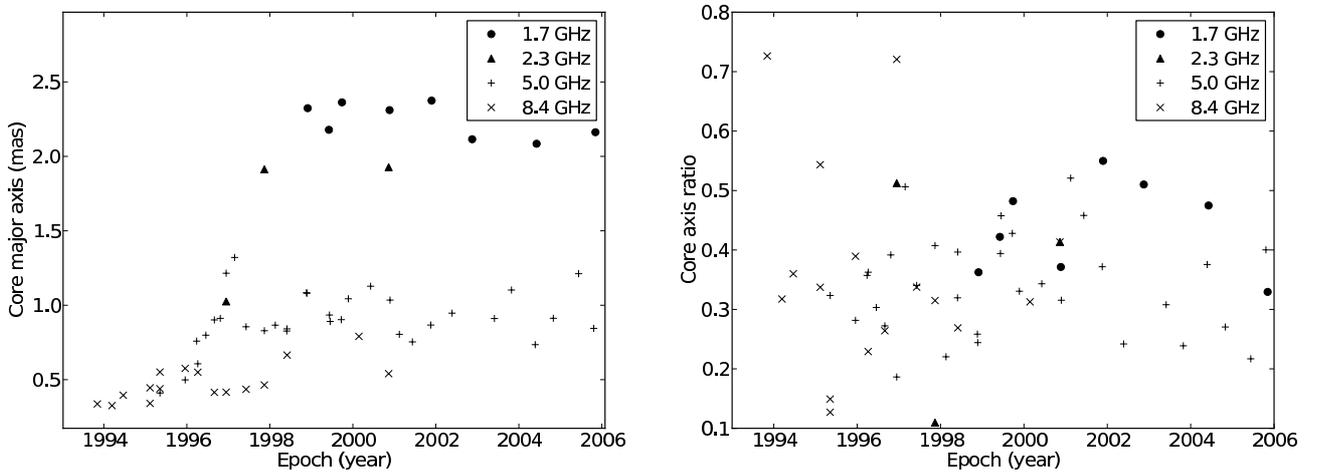}
\caption{Major axis (left) and axis ratio (right), i.e., minor axis in units of major axis, of the 
elliptical Gaussians used to model the core emission of M\,81* at all frequencies (see Table 1 and text).
Typical uncertainties are given in Table \ref{UncertTable}.}
\label{CoreMod2}
\end{figure*}

\begin{figure*}[ht!]
\centering
\includegraphics[width=18cm]{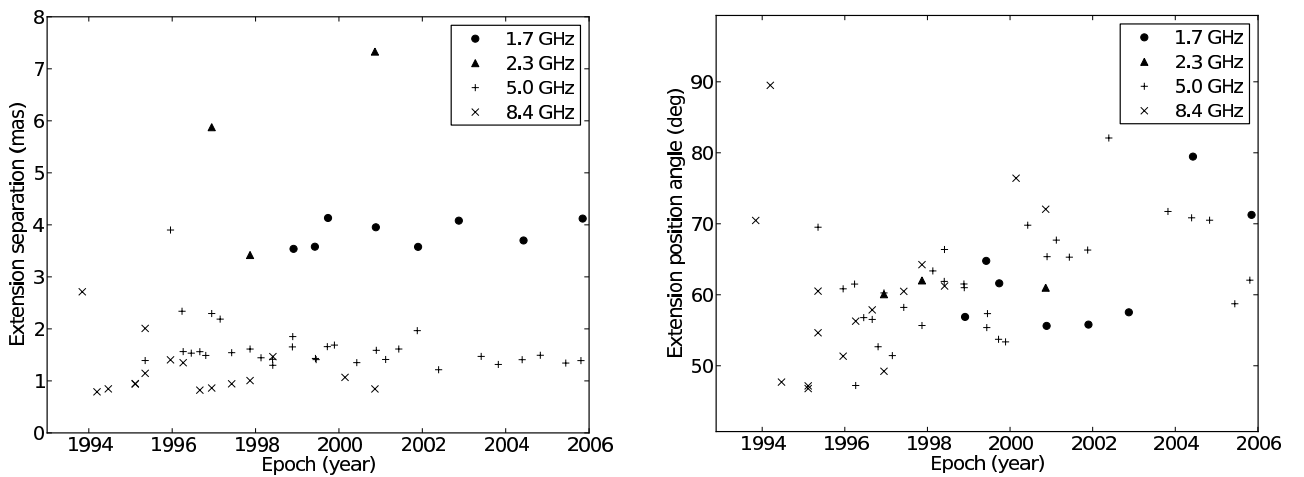}
\caption{Separation from the core (left) and orientation relative to the core (right) of the jet 
extension of M\,81* at all frequencies.}
\label{ExtMod1}
\end{figure*}

\begin{figure}[ht!]
\centering
\includegraphics[width=9cm]{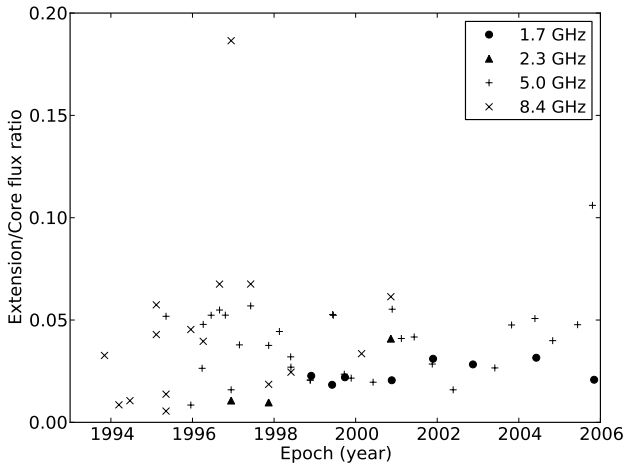}
\caption{Flux density of the M\,81* jet extension relative to the peak intensity of 
the core.}
\label{ExtMod2}
\end{figure}

The defining parameters of the geometrical model described in the previous paragraph were estimated 
for each epoch by model fitting to the visibilities, as implemented in the program {\sc difmap}.
We show the best-fit model parameters at all frequencies and epochs in Table 1 and Figs. \ref{CoreMod1}, 
\ref{CoreMod2}, \ref{ExtMod1}, and \ref{ExtMod2}. 
For the model of the core emission, the parameters are peak intensity, position 
angle (both shown in Fig. \ref{CoreMod1}), major axis, and axis ratio (both shown in Fig. 
\ref{CoreMod2}). For the model of the jet extension, the parameteres are separation from the peak 
of the core, orientation with respect to the core (both shown in Fig. \ref{ExtMod1}), and flux density 
(shown in Fig. \ref{ExtMod2}). We notice that the best-fit
parameters of the core Gaussians are insensitive to the presence of the additional model component used
to describe the jet extension, due to the much smaller contribution of that component to the overall 
structure of the source. (We notice that the flux density of the jet extension is typically below 
7\% of the peak flux density of the core, see Fig. \ref{ExtMod2}.) We have discarded some outliers in Table 1 
to generate Figs. \ref{CoreMod1}, \ref{CoreMod2}, \ref{ExtMod1}, and \ref{ExtMod2}. Basically, we
discarded those fits in which the final axis ratio of the core 
Gaussian was zero (i.e., the peak emission could not be satisfactorily described as a 2-dimensional 
Gaussian) and the few epochs in year 1993 that result in negative position angles.

The typical values for the uncertainties in the model parameters of the core 
Gaussians were estimated using the following approach. For some representative epochs, we found 
the peak in the image of the residuals 
corresponding to the best-fit model. (We call this quantity $\sigma$.) 
Then, we generated a 
synthetic image of the best-fit model, free of noise from the visibilities. We call this image 
$I_{\mathrm{best}}$. Then, to check the range of parameter values that still lead to satisfactory 
fits, we changed the values of the parameters (one parameter at a time), generated the corresponding 
noise-free model images (we call them $I_i$), and computed the maximum deviation between 
$I_{\mathrm{best}}$ and all $I_i$ 
(i.e., $\sigma_i = \mathrm{max}\left(| I_{\mathrm{best}} - I_i |\right) $). The parameter 
uncertainties were estimated as 
the changes in the parameter values such that $\sigma_i = 3\sigma$. Following this approach, 
the resulting uncertainties are those that translate into changes in the model that are three times 
greater than the highest noise contribution to the image. The typical uncertainties 
estimated in this way are given in Table \ref{UncertTable}. These are the uncertainties 
assigned to the parameters at each epoch. However, the uncertainty
in the average of the parameter values for 
a given set of epochs is set to be equal to the standard deviation of 
the parameter values, to reflect the natural scatter in the parameters from epoch to 
epoch.

\begin{table}
\caption{Typical uncertainties in the model parameters of the core Gaussians.}
\centering
\begin{tabular}{c c c c c}
\hline\hline
            & \multicolumn{4}{c}{{\bf Parameter uncertainties}} \\
{\bf Freq.} & PA\tablefootmark{a} & $F_p$\tablefootmark{b} &  Major\tablefootmark{c} & Ratio\tablefootmark{d}    \\
   (GHz)    & (deg) & (mJy\,beam$^{-1}$) &  (mas)  &          \\
\hline
    1.7     &   3   &   0.3            &   0.05  &  0.02    \\
    2.3     &   4   &   4.5            &   0.1   &  0.03    \\
    5.0     &  1.7   &   1.0           &   0.03  &  0.015  \\
    8.4     &  1.5   &   6.5           &  0.04  &  0.02   \\
\hline
\end{tabular}
\\
\tablefoottext{a}{Position angle of the Gaussian.}
\tablefoottext{b}{Peak intensity.}
\tablefoottext{c}{Major axis.}
\tablefoottext{d}{Ratio between minor and major axis.}
\label{UncertTable}
\end{table}

Figure \ref{CoreMod1} (left) shows how the peak intensity at 5.0 and 8.4\,GHz of the fitted Gaussians 
increases from June 1997 to July 1998 and then begins to decrease through June 2001. The maximum peak 
intensity is a factor $\sim$2 of the typical peak intensities outside this four-year-long flare. 
Radio-variabilities of several years have been reported for several 
LLAGN and Seyfert galaxies (Falcke et al. \cite{Falcke2001}; Nagar et al. \cite{Nagar2002}; Mundell 
et al. \cite{Mundell2009}).  However, in this case, as shown below, we also notice hints
of a simultaneous evolution in the orientation of the fitted 
core-Gaussians, which increase the position angle of their 
major axes from $\sim$60 to $\sim$70 degrees (Fig. \ref{CoreMod1} right) during the flare. 
In Sect. \ref{PrecesSec}, we also discuss the possible 
correlation between this flare (together with the changes in the orientation of the fitted core-Gaussians) and our astrometry results at the different frequencies.  

Figure \ref{CoreMod2} shows that the size of the fitted
core-Gaussians (i.e., basically the size of the 
core in the images) is systematically larger at lower frequencies (varies from $\sim$0.5\,mas at 8.4\,GHz to 
$\sim$2\,mas at 1.7\,GHz). The ratio between minor and major Gaussian axes is also higher at lower
frequencies, thus indicating that the core regions at lower frequencies are relatively broader. A larger 
core size at lower frequencies, combined with the frequency-dependent core shift reported in the 
following sections, indicates an opening angle in the jet, as 
discussed in Sect. \ref{OpenAngleSec}.

Figure \ref{ExtMod1} (left) shows how the separation between the peak intensity and the jet extension 
{\em increases} at lower frequencies, with the exception of the data at 2.3 GHz, which gives a larger separation than at 1.7\,GHz. This result is intriguing if we interpret the jet extension as an
optically-thin region in the jet at all frequencies. In such a case, the distance between the central AGN engine
and the (optically-thin) jet extension should be the same at all frequencies. The distance between the core 
and the jet extension should therefore systematically {\em decrease} at lower frequencies if the distance
between the central AGN engine and the core increases (as we show in Sect. \ref{M81jetmodel}). The strange 
behavior seen in Fig. \ref{ExtMod1} (left) may indicate that the jet extensions at different frequencies are spatially 
located in different regions along the jet, although 
other problems related to the very different spatial resolutions achieved at different frequencies (i.e., blending of 
jet components at lower resolutions) could also be affecting the results of our model fitting.

Indeed, we notice that the reported jet extensions at 8.4 and 5 GHz are located within the core at 1.7 GHz (the 
typical core size at 1.7\,GHz is $\sim$2-2.5 mas). Therefore, the jet extension at 1.7 GHz (which is 
located at a distance of $\sim$3-4 mas from the core at the same frequency) must be a physically 
distinct component, indicating that the successive jet extensions at all frequencies are indeed 
tracing different parts of a continuous jet. Figure \ref{ExtMod1} (right) shows that the position angle of 
the jet extension is similar at all frequencies, thus suggesting a rather straight jet in the region 
of the emission at all our observing frequencies. This result is supported in the following 
section, but somewhat in conflict with the results reported in Sect. \ref{PrecesSec}. 

Finally, Fig. \ref{ExtMod2} shows the flux density of the jet 
extension, in units of the core flux density, for all epochs
and frequencies. The typical flux density of the extension is lower than $\sim$7\% of the core flux density, with 
the exception of only two epochs: an epoch at the end of year 1996 (around 10\% of the core flux density at 8.4\,GHz) 
and another at the end of year 2005 (around 18\% of the core flux density at 5.0\,GHz).

\subsection{Extended emission at 1.7\,GHz}
\label{ExtEmisSec}

\begin{figure}[ht!]
\centering
\includegraphics[width=9cm]{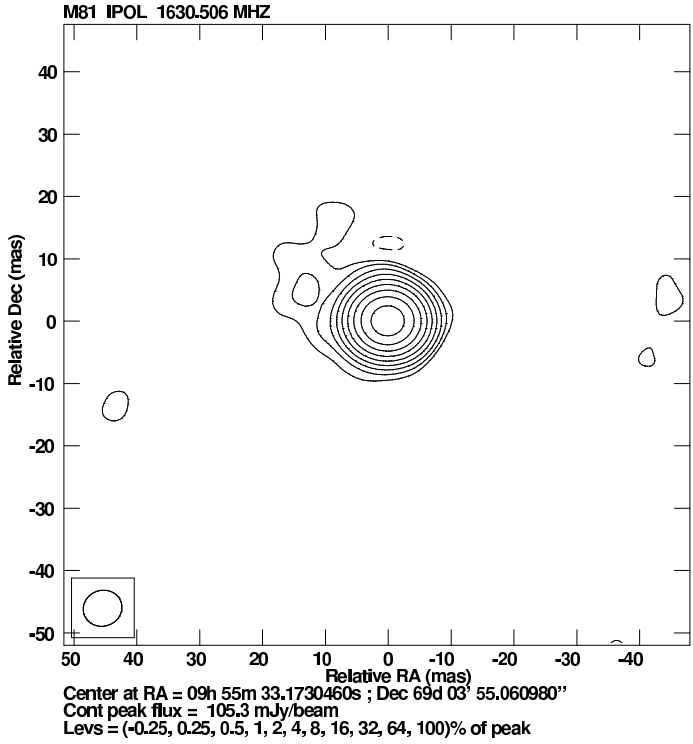}
\caption{Image of M\,81* at 1.7\,GHz obtained from the VLBI observations at epoch 2000 November 20
(see Table 1). The FWHM of the convolving beam is shown in 
the bottom left corner. Natural weighting and a Gaussian taper in Fourier space have been applied
(see text).}
\label{ExtendedFig}
\end{figure}

In the most sensitive low-frequency VLBI observations (those at 1.7\,GHz with large arrays 
and long observing times), we detect the signature of an extended emission centered at $\sim15$\,mas 
from the core with a peak intensity of $\sim$0.7\,mJy\,beam$^{-1}$. The position angle of the 
peak of this extended structure is $\sim65$\,$^{\circ}$ (northeast direction).
We show in Fig. \ref{ExtendedFig} an image with this extended emission. We notice that 
the intensity of this component is so low that natural weighting of the 
visibilities\footnote{Each pixel in the Fourier plane is weighted according to the scatter of 
the visibilities inside it. This weighting scheme improves the sensitivity, but decreases the 
resolution.}, plus an additional downweight of the longest baselines (a Gaussian taper 
with half-width at half-maximum, HWHM, of 30\,M$\lambda$ or lower) is necessary to detect 
the extended emission in the images. Since it is not detected at the higher frequencies, we can 
only set a lower absolute value for the spectral index of this extended component, based on the 
rms of the most sensitive observations at the higher frequencies. We find for this extended 
component that $\alpha < -0.78$ (being the flux density $S \propto \nu^{+\alpha}$).

The similarity of position angles between the jet components reported in the previous
section and this more extended emission points toward a rather straight jet in M\,81* or a jet with 
no strong projected bendings up to a distance of at least $\sim15$\,mas (about 0.26\,pc of projected 
linear scale).

\section{Results for multifrequency astrometry}
\label{ResultsSecII}

\subsection{Frequency-dependent core shifts}
\label{ShiftSec}

\begin{figure}[ht!]
\centering
\includegraphics[width=9.5cm]{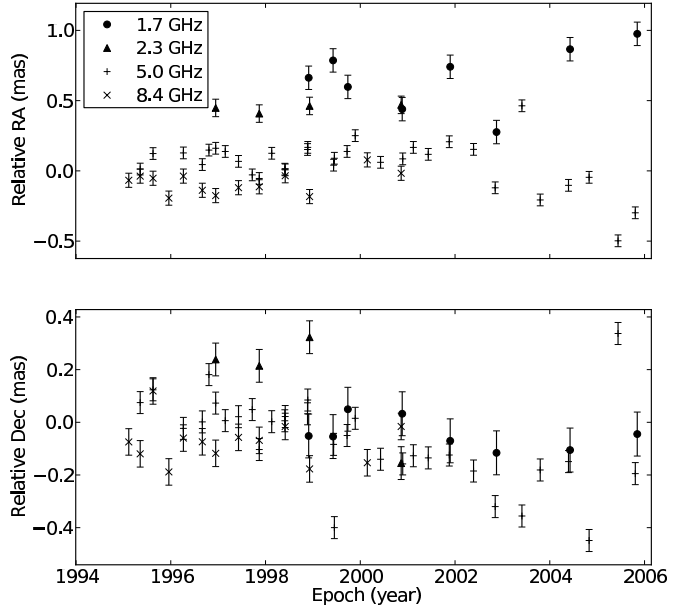}
\caption{Shifts in right ascension and declination of the peak intensity of M\,81*, relative to a 
fixed position on the sky (see text) and using the center of the SN\,1993J radio shell as a 
position reference.}
\label{ProperMotion}
\end{figure}

In Fig. \ref{ProperMotion}, we show the position of the peak of brightness of 
M\,81* relative to the J2000.0 coordinates
$\alpha = 09^{\mathrm{h}}\,55^{\mathrm{m}}\,33.173^{\mathrm{s}}$ and 
$\delta = 69^{\circ}\,3'\,55.062''$, and with the center of the SN\,1993J radio shell 
as a position reference. These coordinates correspond to the nominal position of 
M\,81* used in the correlation of the data at 5\,GHz taken on 30 May 1998.
We set the position of the M\,81* core at this epoch as our coordinate origin. Therefore, 
all the other phase-referenced observations correlated using different nominal positions 
for SN\,1993J and/or M\,81* were accordingly shifted.
In this figure, we show the results for the observations at 8.4, 5.0, 2.3, and 
1.7\,GHz, which are also given in Table 1. The uncertainties are estimated as the half width 
at half maximum (HWHM) of the interferometric beam, divided by the dynamic range of the image 
of SN\,1993J corresponding to each epoch. The uncertainties computed in this way are related 
to the statistical contribution of the image noise in the location of the SN\,1993J radio 
shell. They are also a good estimate of the astrometric precision of compact sources observed
with the phase-referencing technique (e.g., Mart\'i-Vidal et al. \cite{MartiSKA}).

Figure \ref{ProperMotion} shows the dependence of the position of the peak of brightness of M\,81* on
the observing frequency, taking the center of the SN\,1993J shell (at each frequency) as a position
reference. We can use this figure to study the shift of the peak of brightness of M\,81* as a
function of frequency in the context of the relativistic jet model (e.g., Blandford \& K\"onigl
\cite{Blandford}; Lobanov \cite{Lobanov98}). From the time average of the M\,81* positions shown
in Fig. \ref{ProperMotion} at each frequency, we derived the relative shifts shown in Table 
\ref{AstrometryTable} (Row 1) for all pairs of frequencies.

There may be, however, some systematic effects in the data that should be taken into account, or be 
observationally discarded, before we take the shifts in Table \ref{AstrometryTable} (Row 1) as 
true shifts of the brightness peak of M\,81*. On the
one hand, the positions shown in Fig. \ref{ProperMotion} are obtained from
fits of a circularly symmetric shell to the SN\,1993J visibilities. The shell of SN\,1993J does not
have a perfect circular symmetry; there are deviations with respect to a perfect circular source in
both the shape of the shell (of the order of 2\%) and the intensity distribution inside it (of the
order of 20\%) (Bartel et al. \cite{Bartel2002}; Bietenholz, Bartel \& Rupen \cite{BietenPaperIII};
Marcaide et al. \cite{Marcaide2009};
Mart\'i-Vidal et al. \cite{PaperI}). Moreover, the inhomogeneities in the shell evolve
systematically in time. Therefore, and especially for the latest epochs when the supernova
is so extended and weak, there could be unexpected systematics in the estimate of the shell center.
On the other hand, although the proper motion of SN\,1993J in
the galactic coordinates might be small (due to the peculiar motion of the supernova progenitor),
its integrated effect during more than a decade could still leave a small fingerprint in the position
estimates of M\,81* (see the next subsection).

\begin{table*}
\caption{Shifts in the M\,81* brightness peak for several pairs 
of observing frequencies.}
\centering
\begin{tabular}{l | c c c c c c}
\hline\hline
   & \multicolumn{6}{c}{{\bf Frequency pair (GHz)}} \\
           & 1.7--5.0 & 1.7--8.4 & 5.0--8.4 & 1.7--2.3 & 2.3--5.0 & 2.3--8.4 \\
\hline
{\bf Method} & \multicolumn{6}{c}{{\bf Shift (mas)}} \\
Shell center\tablefootmark{a}
 & $0.57\pm0.21$ & $0.74\pm0.19$ & $0.18\pm0.13$  & $0.31\pm0.17$ & $0.42\pm0.14$ & $0.59\pm0.09$ \\
Cross-corr.\tablefootmark{b}
 & $0.68\pm0.22$ & $0.77\pm0.50$ & $0.21\pm 0.10$ & $0.35\pm0.05$ & $0.43\pm0.25$ & $0.65\pm0.14$ \\
\hline
\end{tabular}
\\
\tablefoottext{a}{Taking the estimated center of the SN\,1993J radio shell (at each epoch
and frequency) as position reference.}
\tablefoottext{b}{From the shift in the cross-correlation of SN\,1993J images at different
frequencies and similar epochs.}
\label{AstrometryTable}
\end{table*}

\begin{figure*}[ht!]
\centering
\includegraphics[width=18cm]{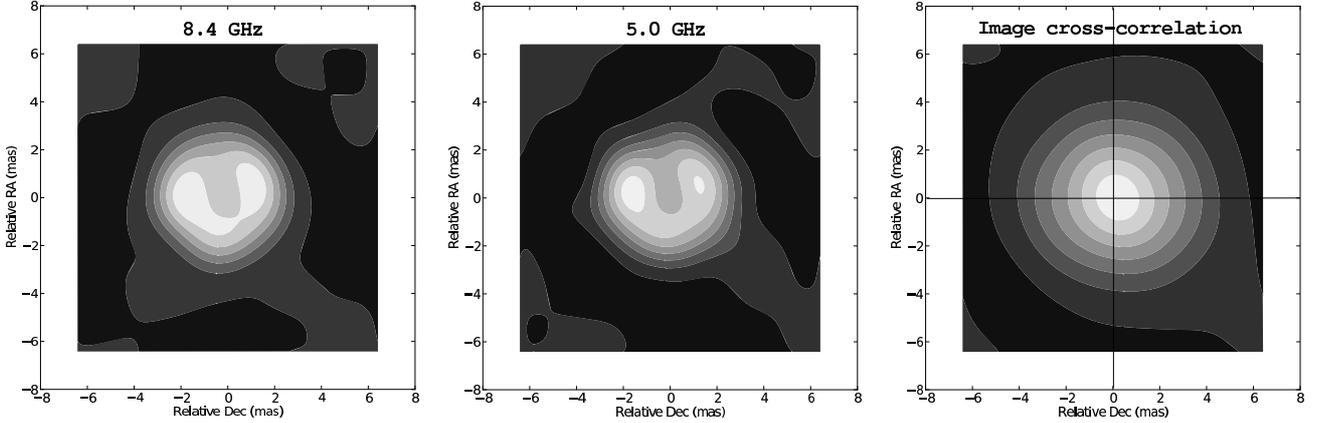}
\caption{Left and center: CLEAN images of SN\,1993J observed at 5.0\,GHz and 8.4\,GHz,
respectively, around September 1996. The color scale runs from black (minimum intensity) to white 
(maximum intensity). Both images are phase-referenced to M\,81*. Right: image
cross-correlation (black=$-0.02$; white=1.00). Notice the small shift in the correlation peak from 
the image origin.}
\label{ImageCorr}
\end{figure*}

We decided to apply a different approach to study the frequency-dependent shift of the M\,81*
brightness peak, which solves, in principle, the drawbacks of using the supernova shell center at
each epoch as a position reference. For the subset of epochs where there are
quasi-simultaneous observations of the supernova at more than one frequency, we convolved the CLEAN
models of SN\,1993J at each epoch with beams of the same size, namely, 0.5
times the shell radius at 5\,GHz reported in Mart\'i-Vidal et al. (\cite{PaperI}). We then
cross-correlated the resulting CLEAN supernova images at different frequencies and, from the peak 
of the cross-correlation image, we estimated the shift of M\,81* between each pair of frequencies. 
In Fig. \ref{ImageCorr}, we show two images of SN\,1993J (at 5 and 8.4\,GHz), corresponding to 
epochs around September 1996, together with their image cross-correlation. The small shift in the 
correlation peak from the center of the image can be appreciated ($\sim0.2$\,mas roughly to the 
west). In Table \ref{AstrometryTable} (Row 2), we present the average shifts of the M\,81* brightness 
peak estimated using this approach. The uncertainties are computed as the standard deviation 
of the averages, to reflect the natural scatter of the data. For all cases, the error bars are 
larger than those determined from model fitting. We used a maximum gap of 15 days between 
two epochs to consider them as quasi-simultaneous\footnote{From the model of Mart\'i-Vidal et al.
(\cite{PaperI}), the expansion of the supernova during 15 days is only $\sim30$\,$\mu$as
on day 500, and less at later times, which is well below the VLBI resolution for all observation frequencies
reported here.} and also required a minimum dynamic range (i.e., peak intensity
over root-mean-square of the residual images) of ten in both SN\,1993J and M\,81*, to use the data of the corresponding epoch.
We notice that the shifts computed from the shell-center estimate (Row 1 in Table \ref{AstrometryTable}),
which are based on model fitting to the visibilities,
are very similar to those computed from the cross-correlation approach (Row 2 of the table), which is
a method completely different from model fitting. Both independent methods therefore give consistent results.

\begin{figure}[ht!]
\centering
\includegraphics[width=9.5cm]{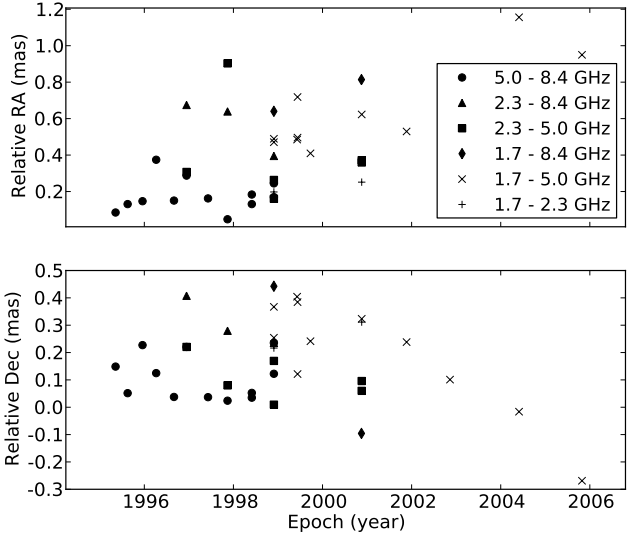}
\caption{Time evolution of the shift in the intensity peak of M\,81*
observed at different pairs of frequencies (peak at the lowest frequency referred to the 
peak at the highest frequency).}
\label{ShiftsFig}
\end{figure}

We show in Fig. \ref{ShiftsFig} the resulting shifts of the M\,81* brightness peak for several pairs
of observing frequencies, obtained with the cross-correlation approach. Any relative motion between SN\,1993J and M\,81* during the
whole observing campaign, and any systematics coming from the fit of a circular shell to the evolving
supernova structure should not affect the results shown in this figure. 
From this figure, there is no clear, unambiguous, systematic time evolution in the shifts during 
the time period covered by our observations, save for the pair of 
frequencies 1.7--5\,GHz at late epochs, with a decrease in relative declination and a 
slight increase in relative right ascension. This evolution could be indicative of precession 
of the intensity peak at 5\,GHz with respect to the peak at 1.7\,GHz. However, these 
latest astrometry results depend on the images of SN\,1993J at late epochs, which are the 
noisiest. The evidence of jet precession resulting only from Fig. \ref{ShiftsFig} should, thus, 
be approached with care. However, in Sect. \ref{PrecesSec} we provide evidence of jet precession resulting
from the combination of the multifrequency astrometry and the morphological evolution of the 
jet.

\subsection{Position stability of M\,81*}
\label{PositionSec}

Figure \ref{ProperMotion} shows a frequency-dependent shift in the brightness peak 
of M\,81*. In Table \ref{PropMotTab}, 
we report the proper motions of those peaks in right ascension, $\mu_{\alpha}$, and declination,
$\mu_{\delta}$, considering all the available data at each frequency. The position 
of the intensity peak at 8.4\,GHz is the most stable one, with a proper motion consistent 
with zero at a level of less than 10\,$\mu$as\,year$^{-1}$. In the other cases but 1.7\,GHz, the proper 
motion fitted in declination is negative and above 3$\sigma$.

The proper motion in right ascension is consistent with zero at all frequencies
but 5.0\,GHz, where a decreasing right ascension of 10\,$\mu$as\,year$^{-1}$ is found at a 
3$\sigma$ level.  
These results differ from those reported in Mart\'i-Vidal et al. 
(\cite{PaperI}), based on the same astrometric data\footnote{We notice, however, that an 
uncertainty of 0.1\,mas was assigned to all the position estimates of the peak intensity of M\,81*
in that publication, which differs from the uncertainties used in the present paper.}, but 
converted into separations and position angles (measured north through east). In that publication, 
we reported that no change in the separation between the intensity peak and our coordinate origin 
was detected at any frequency, with the exception of the data at 5\,GHz beyond July 2001 
(i.e., $\sim$3000 days after the explosion of SN\,1993J). 
This is, indeed, a correct result, but a change in separation consistent with zero was then 
interpreted as a proper motion that is also consistent with zero. Such an interpretation does not
consider the contribution of any possible rotation with respect to 
the reference point, which may contribute, however, to the right-ascension and
declination proper motions shown in Table \ref{PropMotTab}. In either case, the small 
proper motions reported in the present paper translate, at most, into shifts with respect to 
the SN\,1993J shell center of only $1-2$\% of the shell size at any epoch. These shifts are, indeed, 
much smaller than the scatter in the positions of the M\,81* peak, which is similar to the shell-size uncertainties reported in Mart\'i-Vidal et al. (\cite{PaperI}) (see a 
detailed discussion on this matter, and its potential impact in the reported expansion curve 
of SN\,1993J, in the last paragraph of their Sect. 3.1).

\begin{table}
\caption{Proper motion of the intensity peak of M\,81*.}
\centering
\begin{tabular}{l | c c}
\hline\hline
Freq. (GHz) & $\mu_{\alpha}$ (mas\,year$^{-1}$) & $\mu_{\delta}$ (mas\,year$^{-1}$) \\
\hline
1.7 & $0.016\pm0.018$ & $-0.010\pm0.018$ \\
2.3 & $0.008\pm0.022$ & $-0.076\pm0.022$ \\
5.0 & $-0.010\pm0.003$  & $-0.034\pm0.003$ \\
8.4 & $0.000\pm0.007$  & $-0.007\pm0.007$ \\
\hline
\end{tabular}
\label{PropMotTab}
\end{table}

\begin{figure}[ht!]
\centering
\includegraphics[width=9.5cm]{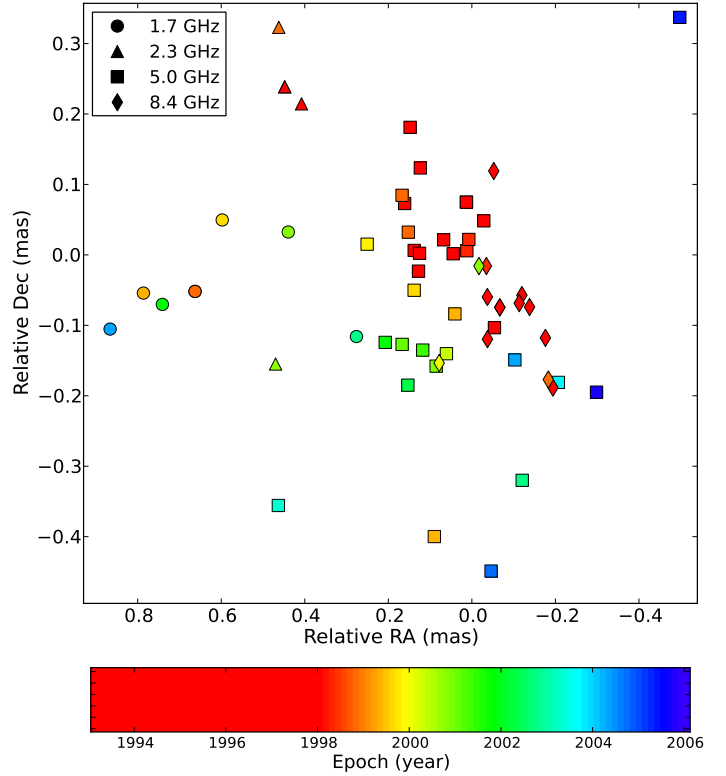}
\caption{Position of the brightness peak of M\,81* at different frequencies (i.e., different symbols)
and different epochs (i.e., different colors). The points plotted are taken from columns $\Delta\alpha$ 
and $\Delta\delta$ of Table 1.}
\label{colorfig}
\end{figure}

As a complementary plot, we show in Fig. \ref{colorfig} the evolution on the sky plane of 
the position of the brightness peak at all frequencies and epochs. On the one hand, different 
frequencies are plotted using different symbols; on the other hand, and for reasons of clarity, the 
results at different epochs are plotted using different colors. Although there are some outsiders 
and the scatter in the data is large, Fig. \ref{colorfig} shows a clear time evolution in the 
overall orientation of the jet, having smaller position angles (i.e., orientation towards north)  
at earlier epochs (i.e., reddish colors) compared to those at later epochs 
(i.e., greenish and blueish colors). The data at 8.4\,GHz are only present at early
epochs (until November 2000), while data at 1.7\,GHz are only present at late epochs (beginning
on November 1998).

\section{Discussion}
\label{DiscussionSec}

\subsection{Spectral-index images}

The radio emission from SN\,1993J at all frequencies comes from the very same location:
the shocked circumstellar region. Therefore, the use of SN\,1993J as a position reference, 
either using the estimated shell center or
cross-correlating the images at different frequencies, allows 
us to build spectral-index images
of M\,81* free of assuming any co-spatial optically-thin jet component at different
frequencies, which is, by far, the most commonly used assumption in the study of chromatic 
effects in AGN jets (e.g., Kovalev et al. \cite{Kovalev}). 

\begin{figure*}[ht!]
\centering
\includegraphics[width=18cm]{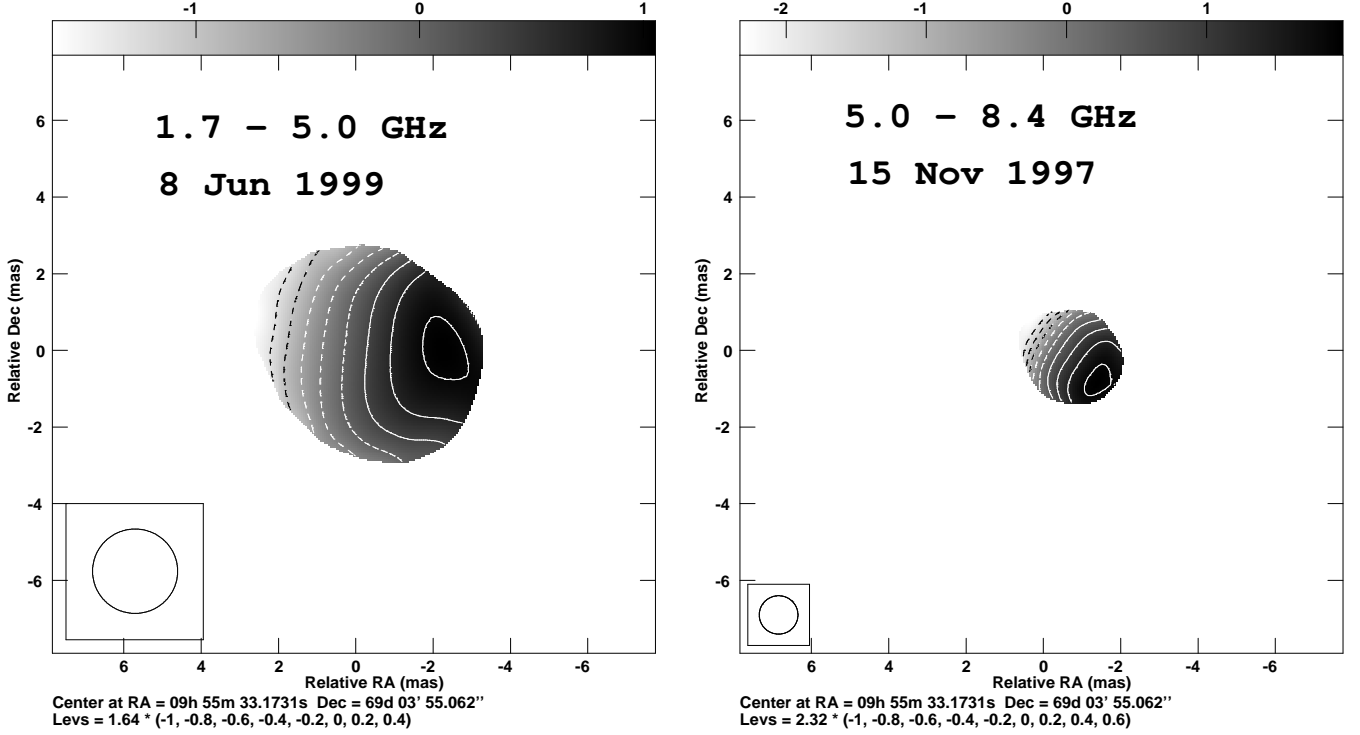}
\caption{Spectral-index images of M\,81* for different pairs of frequencies and epochs.
{\em Epoch} indicates in this case the average date of observation (Table 1, Col. 1) of the 
epochs at both frequencies in each image.}
\label{SpectralFig}
\end{figure*}

We used the shifts in the
SN\,1993J image cross-correlations (see Sect. \ref{ShiftSec}) to align the M\,81* images at 
different epochs and for different pairs of frequencies. The results for two representative 
epochs are shown in Fig. \ref{SpectralFig}. In each image, we convolved the CLEAN models at 
both frequencies using the beam corresponding to the lowest frequency.
To avoid spurious contributions to the spectral-index maps, we clipped those 
points in the image with an intensity lower than 2\% of the peak intensity at each frequency. 
In any case, the spectral indices derived in the regions close to the borders of the source 
should not be taken as reliable, owing to the different sensitivities of the observing arrays at 
different frequencies and the stronger effect of noise in the regions of the source where the 
intensity is low. 

As previously reported in Bietenholz et al. (\cite{Bietenholz2004}), the region around the 
core of M\,81* at each frequency has a flat (or even 
inverted) spectrum, which becomes steeper through the jet as the distance to the core 
increases. For most of the epochs, the gradient of spectral index is aligned with the 
direction of the jet.

This spectral-index gradient along the jet indicates how the opacity is decreasing with
the distance to the jet origin. The same opacity effect indeed explains
the core shift between different observing frequencies (reported in Sect. \ref{ShiftSec}). In the next section, we 
discuss in detail the interpretation of our results in the frame of this jet model.

\subsection{M\,81* and the standard relativistic jet model}
\label{M81jetmodel}

If the magnetic field in the jet and the gradient in the electron 
energy distribution are large enough, synchrotron self-absorption can be relevant.
According to the relativistic jet model, the opacity is changing along the relativistic jet 
and it becomes transparent at a distance from the jet origin that depends on the
observing frequency (Blandford \& K\"onigl \cite{Blandford}). Therefore, it affects the 
apparent position of the AGN jet base (or {\em source core}) at different frequencies.
The dependence of the core position with observing frequency, the {\em core-shift effect} as it 
is known nowadays, was serendipitously discovered 
by Marcaide \& Shapiro (\cite{MarcaideTesis}) and later found in many sources 
(e.g., Kovalev et al. \cite{Kovalev}). If the strength of the magnetic field in the jet, $B$, 
and the particle density, $N$, can both be described as power laws of the distance to the jet 
origin, the core shift, $\Delta r$, between a given pair of observing frequencies, $\nu_1$ and 
$\nu_2$, is given by (Lobanov \cite{Lobanov98})

\begin{equation}
\Delta r = \Omega \left(\frac{\nu_1-\nu_2}{\nu_1\nu_2}\right)^\kappa
\label{ShiftEq}
\end{equation}

\noindent where $\Omega$ (which could be interpreted as the core shift {\em normalized} to
any pair of observing frequencies) depends on the power laws of the decreasing $B$ and $N$. 
Larger $\Omega$ translate into larger shifts of the cores at different frequencies. 
If there is particle-field energy equipartition in the jet, then $\kappa=1$. The value of 
$\Omega$ is the same for any pair of frequencies, $\nu_1$ and $\nu_2$, as long as the 
power laws that characterize the evolution of $B$ and $N$ hold in the jet region where 
the cores at both frequencies are located. 

The average shift in the brightness peak between two frequencies translates into an estimate 
of $\Omega$. We show in Table \ref{ShiftTab} the values of $\Omega$ computed for each pair of 
frequencies shown in Table \ref{AstrometryTable}, Row 1, and assuming $\kappa = 1$. All values of 
$\Omega$ are compatible, although the uncertainties are quite large in all cases (of course, 
the uncertainties are larger for pairs of closer frequencies, since the shifts are smaller 
in these cases; see Eq. \ref{ShiftEq}). The weighted average of $\Omega$ for all the frequency 
pairs is $\Omega = 1.75\pm0.20$\,mas\,GHz.

\begin{table*}
\caption{$\Omega$ values for the jet of M\,81*, derived from different pairs of frequencies.}
\centering
\begin{tabular}{l | c c c c c c}
\hline\hline
 & \multicolumn{6}{c}{{\bf Frequency pair (GHz)}} \\
 &  1.7--5.0 & 1.7--8.4 & 5.0--8.4 & 1.7--2.3 & 2.3--5.0 & 2.3--8.4 \\
\hline
$\Omega$ (mas\,GHz) & $1.49\pm0.55$ & $1.57\pm0.41$ & $2.17\pm1.63$ & $2.04\pm1.09$ & $1.78\pm0.58$ & $1.88\pm0.30$ \\
\hline
\end{tabular}
\label{ShiftTab}
\end{table*}

\subsubsection{Position of the central AGN engine}

Bietenholz, Bartel, \& Rupen (\cite{Bietenholz2000}) have estimated the location of the central engine
of M\,81* based on the most stable position of their geometrical model of M\,81*, referenced to the
shell center of SN\,1993J. Later, Bietenholz, Bartel, \& Rupen (\cite{Bietenholz2004}) followed 
an independent approach to determine the core position, based on 
multifrequency relative astrometry
also referenced to the shell center of SN\,1993J. Both approaches resulted in compatible estimates 
of the location of the central engine in M\,81*.

\begin{table}
\caption{Separation between the jet origin (i.e., $\nu \rightarrow \infty$) and the core observed 
at each frequency.}
\centering
\begin{tabular}{l | c c c c}
\hline\hline
       & \multicolumn{4}{c}{{\bf Frequency (GHz)}} \\
                 & 1.7 & 2.3 & 5.0 & 8.4 \\
\hline
$\Delta r_{\mathrm{core}}$(mas) & $1.02\pm0.10$ & $0.76\pm0.07$ & $0.35\pm0.03$ & $0.21\pm0.02$ \\
\hline
\end{tabular}
\label{ShiftTab2}
\end{table}

From the average value of $\Omega$ reported in the previous section, we can estimate the location 
of jet origin of M\,81* based on a larger dataset than used in Bietenholz, Bartel, \& Rupen 
(\cite{Bietenholz2004}), and using an independent approach (i.e., using Eq. \ref{ShiftEq} to find 
the jet origin). From Eq. \ref{ShiftEq}, and assuming $\kappa = 1$, it can be shown that the 
separation from the jet origin (i.e., the central engine of M\,81*) to the core observed at a given 
frequency $\nu$ is

$$\Delta r_{\mathrm{core}} = \Omega/\nu, $$

\noindent which translates into the separation estimates given in Table \ref{ShiftTab2}. 
The position of the jet origin obtained from our approach (Table \ref{ShiftTab2}) is 
in excellent agreement with the position reported in Bietenholz, Bartel, \& Rupen 
(\cite{Bietenholz2000}, \cite{Bietenholz2004}).

\subsubsection{Magnetic field in the jet and black-hole mass}
 
Since we know the distance to M\,81* (3.63\,Mpc; Freedman et al. \cite{Freedman1994}), 
we can compute the normalized core shift on a linear scale: $\Omega \sim 0.031$\,pc\,GHz. 
This value can be used (together with the distances from the cores to the central engine of the 
AGN, reported in the previous subsection) to estimate the magnetic field in the jet assuming 
particle-field energy equipartition (see, e.g., Eq. 10 in Lobanov \cite{Lobanov98}). For that 
estimate, we also need to know the viewing angle of the jet ($\sim$14 degrees; Devereux et al. 
\cite{Devereux2003}) and its opening angle (a projected value of $\sim$17 
degrees; see Sect. \ref{OpenAngleSec}).

Using Eq. 10 of Lobanov (\cite{Lobanov98}), the estimated magnetic field is 7, 10, 21, and 34\,mG 
in the region of the brightness peak at 1.7, 2.3, 5.0, and 8.4\,GHz, respectively. (The magnetic field
is assumed to be inversely proportional to the distance to the AGN central engine.) Since these are 
just rough estimates, only their order of magnitude is considered in our discussion. 

We notice that these estimates of the magnetic field only depend on the source geometry and the 
differential astrometry (i.e., the core shift). It is intriguing that these estimates are several 
orders of magnitude lower than the magnetic fields derived from the modeling of the M\,81* spectrum 
(tens of Gauss if the inverted spectrum is optically thick; see Reuter \& Lesch 
\cite{Reuter1996}). Much lower magnetic fields can be matched to the spectrum if the 
emission is assumed to be produced by a mono-energetic population of electrons (hence to be 
optically thin; see Reuter \& Lesch \cite{Reuter1996}), but such a scenario is at odds with our 
self-absorption interpretation of the observed core shift.  

Furthermore, as discussed in Lobanov (\cite{Lobanov98}), the magnetic field at 1\,pc 
from the central AGN engine can be used to estimate the mass of the black hole, by direct comparison 
to models of magnetically-driven accretion disks (Field \& Rogers \cite{Field1993}) and black holes 
surrounded by strong magnetic fields (Kardashev \cite{Kardashev1995}). Using this approach, the mass 
of the black hole can be approximated as

$$ M_{\mathrm{bh}} \sim 2.7\times10^9 \,B_1~\mathrm{M}_{\odot}\, $$

\noindent where $B_1$ is the magnetic field at 1\,pc from the AGN engine. With this equation, 
we estimate a mass of $\sim$$2\times10^7$\,M$_{\odot}$. This estimate is within the same order of 
magnitude as the value of $7^{+2}_{-1}\times10^7$\,M$_{\odot}$ reported in Devereux et al. (\cite{Devereux2003}),
based on spectroscopic observations of the central rotating disk of gas, and the value of 
$5.5^{+3.6}_{2.0}\times10^7$\,M$_{\odot}$ reported in Schorr M\"uller et al. (\cite{Schorr2011}), based 
on the stellar-velocity dispersion of the bulge. As a result, our measured 
core shifts, the resulting magnetic field in the jet, and the derived black-hole mass are 
all self-consistent in the frame of the standard jet-interaction model.

\subsection{Jet precession}
\label{PrecesSec}

The model-fitting results shown in Fig. \ref{CoreMod1} (right) hint 
at position angles of the core 
Gaussians that are systematically different at different epochs and 
frequencies. From 1994 to June 2001, the position angle seems to increase from $\sim$50 degrees
to $\sim$70 degrees. (This increase in position angle with time seems to be clearer from June 1997 
onwards, since the data is less spread during the flare from June 1997 to Jun 2001.) 
The change in position angle seems to slow down at later epochs or, even, to 
slightly reverse.

If the major axis of the fitting Gaussian is aligned to the local direction of the jet in the region 
where the peak emission is produced (the most straightforward interpretation of our simplified 
geometrical model), the different position angles at different epochs point to clear evidence of an 
(evolving) bent jet. In the following subsections, we analyze 
the morphological evolution found in M\,81* in more detail 
and discuss its interpretation as due to a precessing jet.

\subsubsection{Evidence of precession from the multifrequency astrometry}

\begin{figure*}[ht!]
\centering
\includegraphics[width=19cm]{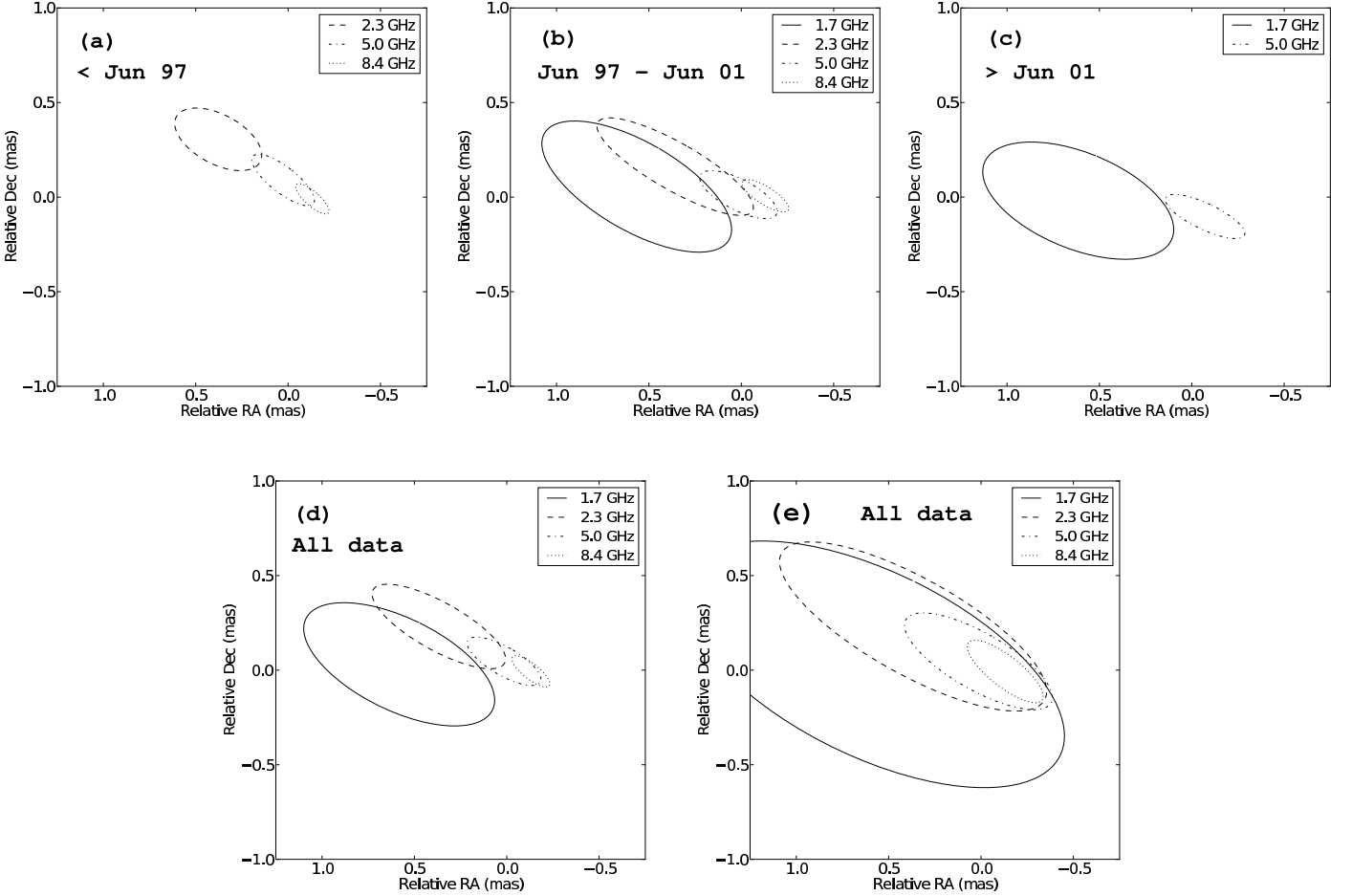}
\caption{Astrometric reconstruction of the jet structure. The ellipses are contours of the Gaussians 
with the parameters given in Table \ref{AvParams}. These parameters are the epoch-average of 
the model-fitting results (Table 1) for different time ranges: (a) up to June 1997; (b) June 1997 through 
June 2001; (c) from June 2001 onwards; and (d) all epochs. In (a)-(d) the contours shown correspond to 0.5 
times the FWHM. (e) is equivalent to (d), but the contours shown correspond to the FWHM. The coordinate 
origin in each figure corresponds to the centroid of the Gaussian at 5\,GHz in (b).}
\label{SchemeFig}
\end{figure*}

We computed the time average of the fitted Gaussian parameters shown in Figs. \ref{CoreMod1} 
and \ref{CoreMod2} for all frequencies and for different time ranges. The results are given in 
Table \ref{AvParams} (the uncertainties are computed as the standard deviations in 
the averages), together with the average, for the {\em same} epochs, of the peak positions reported in 
Sect. \ref{PositionSec}. The quantities given in Table \ref{AvParams} were used to 
generate Fig. \ref{SchemeFig}. Thus, it is a schematic representation of the jet 
structure, based on the
astrometry results reported in Sect. \ref{PositionSec} and the 
model-fitting results reported in Sect. 
\ref{StrucSec}.
For clarity reasons, the contours shown at each frequency correspond to 0.5 times the full width 
at half maximum (FWHM) of the Gaussians (Fig. \ref{SchemeFig} (a)-(d)).

\begin{table*}
\caption{Epoch-average of Gaussian model parameters.}
\centering
\begin{tabular}{l c c c c}
\hline\hline
{\bf Parameter}  & \multicolumn{4}{c}{{\bf Epoch range}} \\
                 & Before June 1997 & June 1997 -- June 2001 & After June 2001 & All data \\
\hline
\multicolumn{5}{c}{{\bf 1.7\,GHz}} \\
$\Delta\alpha$ (mas) &  -- &    $0.64\pm0.12$ & $0.68\pm0.26$  &   $0.65\pm0.18$ \\
$\Delta\delta$ (mas) &  -- &    $-0.01\pm0.05$ & $-0.08\pm0.02$ &   $-0.04\pm0.05$ \\
 PA (deg.)     & --  &    $60.8\pm1.8$   & $68.8\pm3.8$  &  $64.7\pm4.9$ \\
Major (mas)    & --  &    $2.29\pm0.09$  & $2.18\pm0.11$  &  $2.24\pm0.11$ \\
Minor (mas)    & --  &    $0.94\pm0.12$  & $1.02\pm0.21$  &  $0.98\pm0.18$ \\
\hline
\multicolumn{5}{c}{{\bf 2.3\,GHz}} \\
$\Delta\alpha$ (mas) &  $0.45\pm0.05$ &    $0.43\pm0.03$ & --  &  $0.44\pm0.02$ \\
$\Delta\delta$ (mas) &  $0.24\pm0.05$ &    $0.09\pm0.17$ & --  &  $0.16\pm0.15$ \\
 PA (deg.)     &  $62.7\pm4.0$  &    $60.9\pm5.3$   & --  &  $61.5\pm4.4$ \\
Major (mas)    &  $1.03\pm0.10$  &    $1.92\pm0.07$  & --  &  $1.62\pm0.42$ \\
Minor (mas)    &  $0.5\pm0.3$   &    $0.5\pm0.3$  & --  &  $0.51\pm0.24$ \\
\hline
\multicolumn{5}{c}{{\bf 5.0\,GHz}} \\
$\Delta\alpha$ (mas) & $0.09\pm0.05$  &    $0.09\pm0.07$ & $-0.00\pm0.28$  &  $0.08\pm0.10$ \\
$\Delta\delta$ (mas) & $0.02\pm0.06$  &    $-0.05\pm0.11$ & $-0.17\pm0.19$  &  $-0.02\pm0.11$ \\
 PA (deg.)     & $53\pm12$     &    $63.4\pm6.2$   & $65.2\pm4.5$  &  $60.9\pm9.5$ \\
Major (mas)    & $0.82\pm0.29$  &    $0.93\pm0.11$  & $0.94\pm0.14$  &  $0.90\pm0.19$ \\
Minor (mas)    & $0.28\pm0.15$  &    $0.33\pm0.06$  & $0.28\pm0.03$  &  $0.30\pm0.09$ \\
\hline
\multicolumn{5}{c}{{\bf 8.4\,GHz}} \\
$\Delta\alpha$ (mas) &  $-0.06\pm0.06$ &    $-0.06\pm0.07$ & --  &  $-0.06\pm0.06$ \\
$\Delta\delta$ (mas) &  $-0.07\pm0.05$ &    $-0.06\pm0.04$ & --  &  $-0.07\pm0.05$ \\
 PA (deg.)     &  $49.5\pm6.5$  &    $58.5\pm5.7$   & --  &  $52.7\pm7.6$ \\
Major (mas)    &  $0.46\pm0.08$  &    $0.58\pm0.13$  & --  &  $0.50\pm0.12$ \\
Minor (mas)    &  $0.15\pm0.07$  &    $0.19\pm0.04$  & --  &  $0.17\pm0.06$ \\
\hline
\end{tabular}
\\
\tablefoottext{}{The shifts in right
ascension ($\Delta\alpha$) and declination ($\Delta\delta$) refer to the coordinates
given in Sect. \ref{PositionSec}. PA is the position angle of the core Gaussian (north through east 
direction). Major and Minor are the lengths of the Gaussian major and minor axes, 
respectively.}
\label{AvParams}
\end{table*}

A first conclusion from Fig. \ref{SchemeFig} is that, no matter the epoch range 
used in the average, the fitted Gaussians at 1.7\,GHz are always missaligned (towards the southeast) 
with respect to the Gaussians at the other frequencies.
Another conclusion is that the position and inclination of the average Gaussians at different frequencies 
seems to depend on the time range used in the average. Before June 1997, the major axes of the 
Gaussians seem to be well aligned with the overall direction of the jet (Fig. \ref{SchemeFig}(a));
however, between June 1997 and June 2001 (the time range where higher peak flux densities are obtained at 
higher frequencies, see Sect. \ref{StrucSec}), the Gaussians are shifted towards the east. In other words, 
the position angle of the overall jet increases with time. 

This increase in position angle with time is directly related to the decrease of $\Delta\delta$, 
shown in Table \ref{AvParams}, at 1.7, 2.3, and 5.0\,GHz. It must be noticed that the uncertainties in 
$\Delta\delta$ are similar to the changes in $\Delta\delta$ for 
the different epoch ranges used 
in the average, so the increase in the overall position angle of the M\,81* jet seen in
Fig. \ref{SchemeFig} (i.e., taken from Table \ref{AvParams}) might not be statistically 
significant. Nevertheless, we must also notice that a decrease with time in $\Delta\delta$ is observed in 
{\em all cases} at 1.7, 2.3, and 5.0\,GHz (the increase at 8.4\,GHz is only 0.16$\sigma$), 
thus giving more statistical significance to the hypothesis of an increasing
position angle. We elaborate on this by assuming that the {\em null hypothesis} (i.e., a jet 
with a non-evolving position angle) is true. The average in $\Delta\delta$ at all 
frequencies, $\mathrm{Av}(\Delta\delta)$, computed from the results shown in Table \ref{AvParams}, 
is $0.122\pm0.056$\,mas. Therefore,
the {\em Student's ratio} (see, e.g., Mandel \cite{Mandel}) is

$$\frac{\sqrt{n}\, \mathrm{Av}(\Delta\delta)}{\sigma_{\delta}} = 5.32,$$

\noindent where $n$ is the number of pairs of $\Delta\delta$ estimates (6 in our case) 
and $\sigma_\delta$ is the standard deviation of $\Delta\delta$.
The Student's ratio is the deviation (in units of $\sigma$) from the average value predicted 
using the null hypothesis. Higher values of this ratio translate into lower probabilities for the 
null hypothesis. 
If $\Delta\delta$ follows a Gaussian distribution, our set of changes in $\Delta\delta$
follows a Student's distribution, so the likelihood of the null hypothesis can be computed 
from the Student's density function using our estimate of the Student's ratio. The likelihood of the 
null hypothesis (i.e., a nonevolving jet orientation) turns out to be 0.08\%. The likelihood of the 
hypothesis of a jet with an increasing position angle is then 99.92\%. Hence, our astrometry results can 
be considered as strong evidence of an evolving position angle of the jet of M\,81*. The 
fitted proper motions shown in Table \ref{PropMotTab} also point, indeed, to an increasing 
position angle in the jet, since it is oriented in the northeast direction and the proper motions 
in declination are negative at 1.7, 2.3, and 5.0\,GHz. Moreover, from their VSOP observations at 
5\,GHz, Bartel \& Bietenholz (\cite{Bartel2000}) found hints of a helical structure in the M\,81* 
jet, which can be taken as a further support for a changing jet orientation.

Therefore, we find a relation between the increasing position angles of the model core 
Gaussians (Fig. \ref{CoreMod1} right) and the increasing position angle of the overall jet, 
as seen from the differential astrometry at different frequencies (i.e., Fig. \ref{SchemeFig}(a)-(b) 
and Table \ref{AvParams}). These results suggest a changing orientation in the jet of M\,81* 
(e.g., a precessing jet). There is, indeed, a possible additional relation between the changing 
orientation of the jet and the increase in the peak flux density observed at high frequencies 
(Fig. \ref{CoreMod1} left). A jet with an evolving orientation might help for explaining the peak-intensity 
evolution shown in Fig. \ref{CoreMod1} as due, for instance, to changes in the relativistic Doppler 
boosting of the synchrotron emission by the plasma, as it travels 
through the jet, or to the interaction of the jet plasma with inhomogeneities in the surrounding medium
as the jet orientation evolves. In both cases, a periodic evolution in the flux density of M\,81* would 
be expected. 

Indeed, periodic flux-density variability (with periods up to a few years) found in several
AGN have been interpreted as related to jet precession (e.g., the Seyfert galaxy IIIZw2, 
Brunthaler et al. \cite{Brunt2005}; or quasar B0605-085, Kudryavtseva et al.
\cite{Kudriat2011}). In Kudryavtseva et al. (\cite{Kudriat2011}), 
the variability (with a period $7.9\pm0.5$ years)
could be related to the helical path of a jet feature. The period reported for 
B0605-085 is similar to the precession period that we report for M\,81* in Sect. \ref{PASec}. 

It is intriguing that the flux-density flare in M\,81* at 5.0 and 8.4\,GHz is not present at 
1.7\,GHz.  However, there is a 
hint of higher flux densities at 1.7\,GHz before year 2001 (values around 0.11\,Jy) than
after year 2001 (values around 0.08\,Jy; see Fig. \ref{CoreMod1}). This increase in flux density at 
1.7\,GHz is not as strong as the increases at 5.0 and 8.4\,GHz (which are of about a factor 2) and 
could be indicative, for instance, of a smaller Doppler boosting at 1.7\,GHz due to a bent jet (i.e., 
the viewing angle of the part of the jet with maximum emission at 1.7\,GHz would be larger). Indeed, 
the misalignment of the core at 1.7\,GHz with respect to those at the higher frequencies 
(see Fig. \ref{SchemeFig}) might also be related to such a jet bending. This suggests a strong 
bend in the jet orientation within a region of  $\sim0.75-1.0$ mas from the jet origin 
(2600-3600 AU of projected linear scale). Such a bending, however, would conflict with the 
straight jet reported in Sects. \ref{ModFitSec} and \ref{ExtEmisSec}, unless the bending is 
corrected at distances from the jet origin farther than for the core at 1.7\,GHz.

Finally, in Fig. \ref{SchemeFig}(e) we show the average of the fitted models for all epochs, but plotting the
contours of the Gaussians at their FWHM. We can compare this figure directly to Fig. 2 of
Bietenholz et al. (\cite{Bietenholz2004}). The similarity between both figures indicates similar astrometric
results in both works. It must be noticed that the FWHM of the average Gaussians at 2.3, 5.0, and
8.4\,GHz are remarkably well aligned, being the location of the southern side of their major axes
-- associated with the core region at the different frequencies --
almost coincident (as reported in Bietenholz et al. \cite{Bietenholz2004}).

\subsubsection{Evidence of precession from the evolving inclination of the core}
\label{PASec}

\begin{figure}[ht!]
\centering
\includegraphics[width=9.5cm]{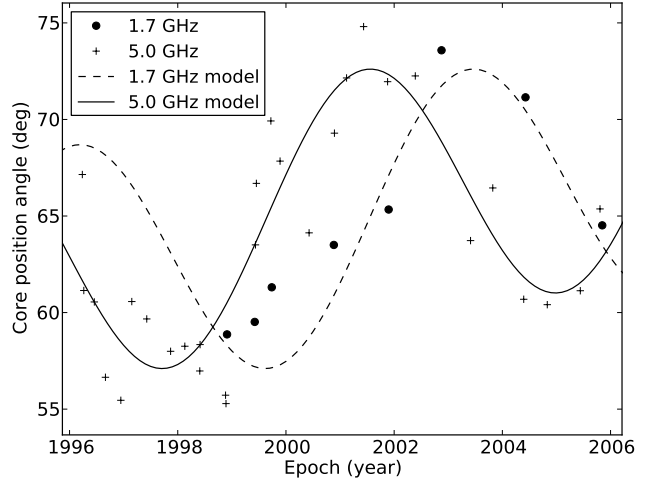}
\caption{Position angle of the core Gaussians at 1.7 and 5\,GHz, observed since year 1996 
(typical uncertainties are given in Table \ref{UncertTable}).
Predictions of our simplified precession model (see Eq. \ref{ModEq} and Table 8) are shown 
in continuous and dashed trace.}
\label{PosAngZoom}
\end{figure}

We can also study the evolution in the position angles of the fitted core Gaussians at each frequency. 
Although the spread in the results shown in Fig. \ref{CoreMod1} (right) is large, 
we notice that the results corresponding to epochs later than year 
1997 are less scattered than those at earlier epochs. This is not surprising, since the overall 
quality of the data and the sensitivity of the interferometric arrays 
(i.e., amount of available radio telescopes, total observing time, and/or 
quality of the receiving systems) was improved year after year. If we focus on the subset of 
epochs at 1.7 and 5\,GHz (i.e., the frequencies with the best array sensitivities) taken after year 
1997, we end up with the 
results shown in Fig. \ref{PosAngZoom}. Although we notice that there is a certain risk of 
overinterpreting these results, an oscillating trend in the position angles at 5\,GHz seems
clear. In the same figure, we show a fitting phenomenological model consisting of a sinusoidal
change of the position angle plus a small drift, i.e.,

\begin{equation}
\theta_\nu = \theta_0 - A \sin{\left(\frac{2\pi}{T} (t - t_\nu) \right)} + \beta (t - t_\nu) ,
\label{ModEq}
\end{equation}

\noindent where $\theta_\nu$ is the position angle of the core Gaussian at frequency $\nu$, 
$t$ the observing epoch, $A$ is the amplitude of oscillation, $T$ the period, $t_\nu$ is 
the reference time at frequency $\nu$, and $\beta$ a small position-angle drift. We 
notice that the only frequency-dependent parameter in this simple model is $t_\nu$. We give 
in Table \ref{ModEqFit} the best-fit values of the model parameters of Eq. \ref{ModEq}. A simple 
interpretation of this model is a precessing jet of period $T$ plus 
a frequency-dependent core shift (which accounts for $t_\nu$). We notice that Fig. 
\ref{PosAngZoom} is compatible with the results shown in Figs. \ref{SchemeFig} and \ref{colorfig} 
in the frame of our jet-precession model, since the increase
in position angle of the core Gaussian at 5\,GHz, owing to the sinusoidal evolution 
shown in Fig. \ref{PosAngZoom}, begins around the end of year 1997 and stops around the 
end of year 2001.

Thus, this increase in
position angle would account for the rotation of the overall jet to the East (Fig. \ref{SchemeFig}(b)).
If our interpretation of Fig. \ref{PosAngZoom} is correct, and the long flare at 5\,GHz and
8.4\,GHz shown in Fig. \ref{CoreMod1} (left) is related to the jet precession, we would
expect to see further similar flares around years 2006 and 2015. 

Similar oscillating trends in the position angles of AGN jets have been found. In particular, 
Lobanov \& Roland (\cite{Lobanov2005}) found an oscillation period of 9.5 years in the inner part 
of the 3C\,345 jet, which was modeled in addition to a long-term drift of 0.4 deg\,year$^{-1}$. 
These results are very similar 
to the fitted parameters shown in Table \ref{ModEqFit}, although the amplitude of oscillation in 
the case of 3C\,345 is around 40 degrees, much larger than that in M\,81*. These authors could 
successfully model their set of 22\,GHz VLBI observations and multiband lightcurves 
with a binary black-hole model plus a precessing accretion disk around the primary.
The period and long-term drift in the evolution of the position angle of the core of M\,81 reported here 
are comparable to those reported by Lobanov \& Roland (\cite{Lobanov2005}) for 3C\,345, but this is 
the first 
time that such behavior has been detected in an LLAGN. In addition, we have been able to relate the 
(multifrequency) changing position angle of the core Gaussians to the (multifrequency) evolving 
core positions in the frame of the host galaxy, via the astrometry referred to SN\,1993J.

\begin{table}
\caption{Best-fit parameters for the simple precession model given in Eq. \ref{ModEq}.}
\centering
\begin{tabular}{c c}
\hline\hline
{\bf Parameter} & {\bf Value} \\
\hline
$\theta_0$ (deg.)            & $62.9\pm0.3$ \\
$A$ (deg)                    & $6.7\pm0.3$  \\
$T$ (years)                  & $7.27\pm0.08$ \\
$\beta$ (deg\,year$^{-1}$)   & $0.54\pm0.07$ \\
$t_{\mathrm{5 GHz}}$          & 1 January 1996 \\
$t_{\mathrm{5 GHz}}-t_{\mathrm{1.7 GHz}}$ (years) & $1.9\pm0.4$ \\
\hline
\end{tabular}
\label{ModEqFit}
\end{table}

\subsection{Opening angle of the jet and precession tilt}
\label{OpenAngleSec}

Another conclusion that can be extracted from Fig. \ref{SchemeFig} is that the jet becomes progressively wider as 
the distance to the core increases, indicating a conic-like section in its longitudinal direction. The (projected) 
opening angle of the jet, estimated from the contours shown in Fig. \ref{SchemeFig}(a), is 
$\sim$17 degrees. Since this angle has been derived from the projection of the jet on the sky plane, 
it is an upper limit to the true opening angle of the jet. 

If we assume that the jet is almost perpendicular to the plane of the rotating parsec-scale disc of 
gas reported in Devereux et al. (\cite{Devereux2003}), the jet viewing angle is $\sim14$ degrees. 
Hence, the true opening angle of the jet would be $\sim4$ degrees.

In the scenario of a precessing jet, the angle between the precession axis and the jet (i.e., the 
tilt angle of the jet) would be around a half of the total change in position angle of the 
major axes of the fitted core Gaussians (which is $\sim$7 degrees, see the value of 
parameter $A$ in Table \ref{ModEqFit}). Therefore, the real tilt angle in the precession 
(i.e., correcting for the jet viewing angle) would be $\sim$1 degree.

\section{Conclusions}
\label{ConclusionsSec}

We have reported on a set of global VLBI observations of the AGN in M81 (M\,81*), performed 
between years 1993 and 2005 at the frequencies of 1.7, 2.3, 5.0, and 8.4\,GHz. We 
studied the morphological evolution of the source and its differential astrometry with 
respect to SN\,1993J (a stable position reference in the galactic frame of M\,81).

The source consists, at all frequencies, of a slightly resolved core component plus a 
weak jet extension in the northeast direction (position angle of about 65 degrees). 
The core component has an increasing size with decreasing frequency (from $\sim0.5$\,mas 
at 8.4\,GHz to $\sim2.2$\,mas at 1.7\,GHz). At 1.7\,GHz we also find a more extended emission  
located at$\sim15$\,mas from the core and at a position 
angle of $\sim65$ degrees. This extended emission hints at a rather straight jet in 
M\,81* at least up to 0.26\,pc (1.1\,pc) of projected (unprojected) distance to the 
core. The core components at all frequencies are elongated roughly in the direction of the 
jet extension and show evidence of an increasing 
position angle. 

The evolution of the peak flux density of the core shows a clear long-term trend at 5.0 
and 8.4\,GHz, with an increase starting around June 1997, a maximum reached 
around July 1998, and a decrease ending around June 2001. The flux-density variability 
during this long-term flare is about 100\% with respect to that in the 
quiescent stage. 

For the astrometric study, we find a shift in the brightness 
peak of M\,81* towards the southwest direction with increasing frequency, 
as previously reported in Bietenholz, Bartel, \& Rupen (\cite{Bietenholz2004}). We 
compared these shifts with the predictions of the relativistic jet model, assuming 
a power-law behavior for the decay in the magnetic field and the particle density 
distribution along the jet. 
We used the Lobanov (\cite{Lobanov98}) model to estimate the position of the central 
engine of M\,81* (i.e., the origin of the jet), relative to the average position of the 
brightness peaks of our VLBI images. The estimated position of the jet origin is in 
excellent agreement with the one reported in previous publications using different 
approaches (Bietenholz, Bartel, \& Rupen \cite{Bietenholz2000}, \cite{Bietenholz2004}). 
We also used the model of Lobanov (\cite{Lobanov98}) to estimate the magnetic field
in the jet and the mass of the central black hole. We obtained magnetic fields of a few milli-Gauss at distances of a fraction of a parsec from the central engine of the 
AGN. These values are much lower than those needed to fit the (self-absorption dominated) 
inverted spectrum of M\,81*. However, we obtained a black-hole mass of $\sim$$2\times10^7$\,$M_{\odot}$, 
comparable to previously reported estimates using different approaches.

From the spectral-index images of M\,81*, we find a region with inverted spectrum (typical
values of $\alpha$ between 0.5 and 1.5) around the intensity peak at each frequency. The
spectrum becomes steeper along the jet. This behavior, together with the multiwavelength
astrometric results, agrees with the relativistic jet model.

The position of the peak intensity at 8.4\,GHz is very stable in the frame of the host 
galaxy, with a proper motion consistent with zero to better than 10\,$\mu$as\,year$^{-1}$. 
The positions of the peaks at the other frequencies are less stable, and they systematically 
evolve towards the south (i.e., larger position angles) at a rate of 10, 76, and
34\,$\mu$as\,year$^{-1}$ at 1.7, 2.3, and 5.0\,GHz, respectively. We interpret these 
results as strong evidence of jet precession towards the south. 

There is also a relation between the evolution in the relative astrometry among the 
peaks at different frequencies and the evolving inclination of the cores, as fitted with 
elliptical Gaussians. This relation lends further support
to an evolving orientation of the overall jet of M\,81*. The long-term variability found in
the peak flux densities at 5.0 and 8.4\,GHz also seems related to the evolving
jet orientation. In such a case, if the changing orientation of the jet is periodic (i.e., 
the jet is precessing, so that its position angle will eventually turn northward after reaching 
its maximum value, around 70 degrees, and then turn back eastward), we predict that the flux 
density of M\,81* at the higher frequencies will eventually increase and follow a 
light curve similar to our observations by the next 10 years. 
Thus, a 
continued monitoring of the flux density and jet structure evolution of this LLAGN will be 
necessary to further confirm (and, eventually, better constrain) our precession model.

An increase in the accretion rate of the central engine and its later correlation with the 
jet particle density seem unlikely to explain the light curve of Fig. \ref{CoreMod1} (left), since it would 
not explain the simultaneous change in the jet orientation. 
Therefore, one would be tempted to 
think of a correlation between the flare and the evolving jet orientation, which would, in principle,
imply the periodic generation of similar long-term flares in the future, as in the case of 
B0605-085 (Kudryavtseva et al. \cite{Kudriat2011}).

\begin{acknowledgements}

IMV is a fellow of the Alexander von Humboldt foundation in Germany.
MAPT acknowledges support by the Spanish Ministry of Education and
Science (MEC), through grant AYA2006-14986-C02-01, and by the Consejer\'ia
de Innovaci\'on, Ciencia y Empresa of Junta de Andaluc\'ia, through grants
FQM-1747 and TIC-126.
Partial support from Spanish grants AYA2009-13036-C02-02, 
AYA2009-13036-C02-02, and Prometeo 2009/104 is also acknowledged. The 
National Radio Astronomy 
Observatory is a facility of the National Science Foundation 
operated under cooperative agreement by Associated Universities, 
Inc. The European VLBI Network is a joint facility of European, 
Chinese, South African, and other radio astronomy institutes
funded by their national research councils.

\end{acknowledgements}

\onecolumn
\longtab{1}{
\renewcommand{\thefootnote}{\alph{footnote}}
\begin{longtable}{ l c c c c c c c }
\caption{Results of astrometry and model fitting of M\,81*. In the cases when it 
was not possible to perform a reliable astrometry (due to poor uv-coverage, too 
noisy data, or both), we added ``--''.}\\

\hline\hline
Epoch & $\Delta\alpha$\footnotemark[1] & $\Delta\delta$\footnotemark[1] & $\sigma$\footnotemark[1] 
& Peak\footnotemark[2] & PA\footnotemark[2] & Major\footnotemark[2] & Minor\footnotemark[2] \\
(dd/mm/yy) & (mas) & (mas) & (mas) &(mJy\,beam$^{-1}$) & (deg) & (mas) & (mas) \\
\hline

\endfirsthead

\caption{continued.}\\

\hline\hline
Epoch & $\Delta\alpha$\footnotemark[1] & $\Delta\delta$\footnotemark[1] & $\sigma$\footnotemark[1] 
& Peak\footnotemark[2] & PA\footnotemark[2] & Major\footnotemark[2] & Minor\footnotemark[2] \\
(dd/mm/yr) & (mas) & (mas) & (mas) & (mJy\,beam$^{-1}$) & (deg) & (mas) & (mas) \\
\hline
\endhead

\hline
\endfoot

 \multicolumn{8}{c}{{\bf 1.7\,GHz}} \\
06/11/93  &  --  &  --  &  --  &  0.086  &  $-47.0$  &  1.552  &  0.637  \\
17/12/93  &  --  &  --  &  --  &  0.055  &  42.0  &  4.293  &  0.0  \\
15/03/94  &  --  &  --  &  --  &  0.05  &  23.7  &  4.225  &  0.0  \\
30/11/98  &  0.663  &  $-0.052$  &  0.09  &  0.106  &  58.9  &  2.324  &  0.842  \\
06/06/99  &  0.786  &  $-0.054$  &  0.09  &  0.118  &  59.5  &  2.178  &  0.919  \\
28/09/99  &  0.598  &  0.050  &  0.09  &  0.116  &  61.3  &  2.363  &  1.139  \\
20/11/00  &  0.440  &  0.030  &  0.10  &  0.115  &  63.5  &  2.311  &  0.858  \\
26/11/01  &  0.741  &  $-0.07$  &  0.11  &  0.073  &  65.3  &  2.375  &  1.306  \\
17/11/02  &  0.276  &  $-0.116$  &  0.12  &  0.085  &  73.6  &  2.115  &  1.079  \\
06/06/04  &  0.866  &  $-0.105$  &  0.14  &  0.079  &  71.1  &  2.085  &  0.990  \\
06/11/05  &  0.975  &  $-0.045$  &  0.15  &  0.080  &  64.5  &  2.162  &  0.712  \\
\hline
 \multicolumn{8}{c}{{\bf 2.3\,GHz}} \\
19/12/95  &  --  &  --  &  --  &  0.092  &  50.4  &  1.512  &  0.0  \\
13/12/96  &  0.448  &  0.239  &  0.050  &  0.103  &  62.7  &  1.026  &  0.526  \\
15/11/97  &  0.408  &  0.214  &  0.050  &  0.098  &  55.6  &  1.913  &  0.211  \\
07/12/98  & -- & -- & -- & -- & -- & -- & -- \\
13/11/00  &  0.47  &  $-0.155$  &  0.080  &  0.095  &  66.3  &  1.927  &  0.797  \\
\hline
 \multicolumn{8}{c}{{\bf 5.0\,GHz}} \\
06/11/93  &  --  &  --  &  --  &  0.101  &  26.0  &  0.787  &  0.0  \\
17/12/93  &  --  &  --  &  --  &  0.080  &  78.3  &  5.224  &  0.0  \\
28/01/94  &  --  &  --  &  --  &  0.087  &  75.1  &  1.541  &  0.0  \\
15/03/94  &  --  &  --  &  --  &  0.065  &  11.8  &  0.021  &  0.0  \\
22/04/94  &  --  &  --  &  --  &  0.086  &  45.8  &  1.148  &  0.0  \\
22/06/94  &  --  &  --  &  --  &  0.095  &  $-34.0$  &  0.873  &  0.165  \\
30/08/94  &  --  &  --  &  --  &  0.060  &  73.2  &  2.089  &  0.0  \\
31/10/94  &  --  &  --  &  --  &  0.093  &  26.7  &  0.582  &  0.0  \\
23/12/94  &  --  &  --  &  --  &  0.082  &  10.7  &  0.472  &  0.0  \\
11/05/95  &  --  &  --  &  --  &  0.095  &  26.0  &  0.41  &  0.133  \\
18/08/95  &  0.122  &  0.123  &  0.023  &  0.101  &  44.9  &  0.717  &  0.0  \\
19/12/95  &  0.126  &  0.047  &  0.025  &  0.100  &  41.1  &  0.499  &  0.140  \\
28/03/96  &  0.128  &  $-0.016$  &  0.03  &  0.110  &  67.1  &  0.759  &  0.271  \\
08/04/96  &  0.128  & $-0.022$  &  0.03  &  0.106  &  61.1  &  0.607  &  0.220  \\
17/06/96  &  0.088  &  $-0.011$  &  0.03  &  0.104  &  60.6  &  0.799  &  0.242  \\
01/09/96  &  0.045  &  0.001  &  0.03  &  0.116  &  56.6  &  0.901  &  0.245  \\
22/10/96  &  0.148  &  0.181  &  0.03  &  0.096  &  46.6  &  0.912  &  0.357  \\
13/12/96  &  0.161  &  0.073  &  0.03  &  0.136  &  55.5  &  1.215  &  0.226  \\
25/02/97  &  0.138  &  0.006  &  0.03  &  0.105  &  60.6  &  1.321  &  0.669  \\
07/06/97  &  0.068  &  0.021  &  0.03  &  0.113  &  59.7  &  0.855  &  0.291  \\
15/11/97  &  $-0.054$  &  $-0.103$  &  0.04  &  0.105  &  58.0  &  0.829  &  0.338  \\
18/02/98  &  0.125  &  0.002  &  0.04  &  0.123  &  58.3  &  0.866  &  0.191  \\
30/05/98  &  0.012  &  0.006  &  0.04  &  0.191  &  57.0  &  0.827  &  0.264  \\
03/06/98  &  0.009  &  0.017  &  0.04  &  0.154  &  58.3  &  0.841  &  0.334  \\
20/11/98  &  0.151  &  0.032  &  0.04  &  0.224  &  55.7  &  1.084  &  0.28  \\
23/11/98  &  0.167  &  0.085  &  0.04  &  0.229  &  55.3  &  1.08  &  0.264  \\
10/06/99  &  0.041  &  $-0.084$  &  0.05  &  0.144  &  63.5  &  0.934  &  0.368  \\
16/06/99  &  0.090  &  $-0.400$  &  0.05  &  0.145  &  66.7  &  0.891  &  0.408  \\
22/09/99  &  0.138  &  $-0.05$  &  0.05  &  0.202  &  69.9  &  0.903  &  0.386  \\
24/11/99  &  0.249  &  0.014  &  0.05  &  0.185  &  67.8  &  1.043  &  0.345  \\
06/06/00  &  0.061  &  $-0.140$  &  0.05  &  0.169  &  64.1  &  1.128  &  0.387  \\
24/11/00  &  0.085  &  $-0.158$  &  0.05  &  0.133  &  69.3  &  1.034  &  0.326  \\
14/02/01  &  0.167  &  $-0.127$  &  0.06  &  0.106  &  72.1  &  0.805  &  0.419  \\
10/06/01  &  0.118  &  $-0.135$  &  0.06  &  0.121  &  74.8  &  0.754  &  0.345  \\
18/11/01  &  0.207  &  $-0.124$  &  0.06  &  0.097  &  72.0  &  0.867  &  0.322  \\
24/05/22  &  0.153  &  $-0.185$  &  0.06  &  0.123  &  72.3  &  0.947  &  0.229  \\
01/06/03  &  0.463  &  $-0.356$  &  0.07  &  0.121  &  63.7  &  0.911  &  0.280  \\
29/03/03  &  $-0.203$  &  $-0.180$  &  0.07  &  0.068  &  66.5  &  1.102  &  0.263  \\
25/05/04  &  $-0.103$  &  $-0.149$  &  0.07  &  0.098  &  60.7  &  0.735  &  0.276  \\
31/10/04  &  $-0.046$  &  $-0.449$  &  0.07  &  0.105  &  60.4  &  0.912  &  0.247  \\
11/06/05  &  $-0.498$  &  0.337  &  0.08  &  0.111  &  61.1  &  1.212  &  0.263  \\
22/10/05  &  $-0.298$  &  $-0.195$  &  0.08  &  0.107  &  65.4  &  0.845  &  0.338  \\
\hline
 \multicolumn{8}{c}{{\bf 8.4\,GHz}} \\
04/08/93  &  --  &  --  &  --  &  0.101  &  $-71.2$  &  1.233  &  0.655  \\
19/09/93  &  --  &  --  &  --  &  0.069  &  $-6.9$  &  0.291  &  0.0  \\
06/11/93  &  --  &  --  &  --  &  0.022  &  52.0  &  0.337  &  0.245  \\
28/01/94  &  --  &  --  &  --  &  0.050  &  37.1  &  0.799  &  0.0  \\
15/03/94  &  --  &  --  &  --  &  0.095  &  39.0  &  0.326  &  0.103  \\
22/06/94  &  --  &  --  &  --  &  0.098  &  51.2  &  0.395  &  0.142  \\
12/02/95  &  $-0.067$  &  $-0.075$  &  0.02  &  0.110  &  56.2  &  0.341  &  0.185  \\
11/05/95  &  $-0.038$  &  $-0.119$  &  0.03  &  0.101  &  49.3  &  0.439  &  0.056  \\
18/08/95  &  $-0.052$  &  0.117  &  0.03  &  0.098  &  48.8  &  0.321  &  0.0  \\
19/12/95  &  $-0.194$  &  $-0.188$  &  0.03  &  0.116  &  51.7  &  0.576  &  0.224  \\
08/04/96  &  $-0.039$  &  $-0.061$  &  0.03  &  0.110  &  53.7  &  0.55  &  0.126  \\
01/09/96  &  $-0.137$  &  $-0.074$  &  0.04  &  0.112  &  53.5  &  0.415  &  0.110  \\
13/12/96  &  $-0.176$  &  $-0.118$  &  0.04  &  0.133  &  32.2  &  0.416  &  0.301  \\
07/06/97  &  $-0.120$  &  $-0.058$  &  0.04  &  0.109  &  51.2  &  0.435  &  0.147  \\
15/11/97  &  $-0.113$  &  $-0.069$  &  0.05  &  0.158  &  54.7  &  0.464  &  0.146  \\
03/06/98  &  $-0.035$  &  $-0.016$  &  0.05  &  0.205  &  56.6  &  0.665  &  0.179  \\
25/02/00  &  0.078  &  $-0.154$  &  0.06  &  0.193  &  63.3  &  0.791  &  0.247  \\
13/11/00  &  $-0.017$  &  $-0.016$  &  0.06  &  0.146  &  66.7  &  0.54  &  0.223  \\

\footnotetext[1]{$\Delta\alpha$ and $\Delta\delta$ are the right ascension and declination 
(respectively) of the intensity peak referred to the coordinates given in Sect. 
\ref{PositionSec}. $\sigma$ is the uncertainty in both, $\Delta\alpha$ and $\Delta\delta$.}
\footnotetext[2]{Peak, PA, Major, and Minor are the intensity peak,
position angle, major axis, and minor axis (respectively) of Gaussians fitted
to the M\,81* visibilities. }
\end{longtable}
\label{AllObs}
}


\begin{thebibliography}{.99}


	
\bibitem[1986]{Alef1986} Alef W. \& Porcas R.W. 1986, A\&A, 168, 365

\bibitem[2000]{Bartel2000} Bartel, N. \& Bietenholz, M.F. 2000, in ``Proceedings of 
                           the VSOP Symposium'', eds. H. Hirabayashi, P.G. Edwards, 
                           and D.W. Murphy, (Sagamihara: Institute of Space and 
                           Astronautical Science), 17

\bibitem[2002]{Bartel2002} Bartel, N., Bietenholz, M.~F., Rupen, M.~P., et al.
                           2002, ApJ, 581, 404

\bibitem[2000]{Bietenholz2000} Bietenholz, M.~F., Bartel, N., \& Rupen, M.~P.
                               2000, ApJ, 532, 895

\bibitem[2000]{BietenPaperIII} Bietenholz, M.~F., Bartel, N., \& Rupen, M.~P.
                               2003, ApJ, 597, 374

\bibitem[2004]{Bietenholz2004} Bietenholz, M.~F., Bartel, N., \& Rupen, M.~P.
                               2004, ApJ, 615, 173

\bibitem[1979]{Blandford} Blandford, R.~D. \& Konigl, A. 1979, ApJ, 232, 34

\bibitem[2005]{Brunt2005} Brunthaler, A., Falcke, H., Bower, G.~C., et al. 2005,
                          A\&A 435, 497 
\bibitem[2003]{Devereux2003} Devereux, N., Ford, H., Tsvetanov, Z., \& Jacoby, G. 
                             2003, AJ, 125, 1226

\bibitem[1994]{Duschl1994} Duschl, W.J., \& Lesch, H. 1994, A\&A, 286, 431

\bibitem[2001]{Falcke2001} Falcke, H., Leh{\'a}r, J., Barvainis, R., Nagar, N.M., 
\& Wilson, A.S.\ 2001, in ``Probing the Physics of Active Galactic Nuclei'',
ASP Conference Proceedings Vol. 224, 265 

\bibitem[1993]{Field1993} Field, G.B. \& Rogers, R.D. 1993, ApJ, 403, 94

\bibitem[1994]{Freedman1994} Freedman, W.L., Hughes, S.M., Madore, B.F., 
                             et al. 1994, ApJ, 427, 628

\bibitem[1999]{Ho1996} Ho, L.C., Filippenko, A.V., \& Sargent, W.L.W. 1996, 
                       ApJ, 462, 183

\bibitem[1999]{Ho1999} Ho, L.C., van Dyk, S.D., Pooley, G.G., et al. 1999, AJ, 118, 843

\bibitem[1996]{Ishisaki1996} Ishisaki, Y., Makishima, K., Iyomoto, N., et al. 
                             1996, PASJ, 48, 237

\bibitem[1995]{Kardashev1995} Kardashev, N.S. 1995, MNRAS, 276, 515

\bibitem[2006]{Kettenis2006} Kettenis, M., van Langevelde, H. J., Reynolds, C.,
\& Cotton, B. 2006, Astronomical Data Analysis Software and Systems XV, 351, 497

\bibitem[2008]{Kovalev} Kovalev, Y.~Y., Lobanov, A.~P., Pushkarev, A.~B.,
                        \& Zensus, J.~A.\ 2008, A\&A, 483, 759

\bibitem[2011]{Kudriat2011} Kudryavtseva, N.A., Britzen, S., Witzel, A., et al. 2011,
                            A\&A, 526A, 51

\bibitem[1998]{Lobanov98} Lobanov, A.P. 1998, A\&A, 330, 79
	
\bibitem[2005]{Lobanov2005} Lobanov, A.P. \& Roland, J. 2005, A\&A, 431, 831

\bibitem[1964]{Mandel} Mandel, J. 1964, ``The Statistical Analysis of Experimental
                       Data'', Dover Publications Inc., New York

\bibitem[1984]{MarcaideTesis} Marcaide, J.M. \& Shapiro, I.~I. 1984,
                              ApJ, 276, 56

\bibitem[2009]{Marcaide2009} Marcaide, J.~M., Mart\'i-Vidal, I., Alberdi, A.,
                             et al. 2009, A\&A, 505, 927

\bibitem[2008]{Markoff2008} Markoff, S., Nowak, M., Young, A., et al. 2008, ApJ, 681, 905

\bibitem[2002]{Marscher2002} Marscher, A.P., Jorstad, S., G\'omez, J.L., et al. 2002, Nat, 417, 625

\bibitem[2010]{MartiSKA} Mart{\'{\i}}-Vidal, I., Guirado, J.C., Jim{\'e}nez-Monferrer, S., 
                     \& Marcaide, J.M. 2010, A\&A, 517, A70

\bibitem[2011a]{PaperI} Mart\'i-Vidal, I., Marcaide, J.M., Alberdi, A., et al.
                        2011, A\&A, 526A, 142

\bibitem[2011b]{PaperII}  Mart\'i-Vidal, I., Marcaide, J.M., Alberdi, A., et al.
                        2011, A\&A, 526A, 143

\bibitem[2003]{Merloni2003} Merloni, A., Heinz, S., \& Di Matteo, T. 2003, MNRAS, 345, 1057

\bibitem[1998]{Mirabel1998} Mirabel, I.F., Rodr\'{\i}guez, L.F. 1998, Nat, 392, 673

\bibitem[2010]{Miller2010} Miller, J.M., Nowak, M., Markoff, S., et al. 2010, ApJ, 720, 1033

\bibitem[2009]{Mundell2009} Mundell, C.G., Ferruit, P., Nagar, N., \& Wilson, A.S.\ 2009, 
                            ApJ, 703, 802 

\bibitem[2002]{Nagar2002} Nagar, N.M., Falcke, H., Wilson A.S., \& Ulvestad, J.S. 2002, 
                          A\&A, 392, 53

\bibitem[1994]{Page2004} Page, M.J., Soria, R., Zane, et al. 2004, A\&A, 422, 77

\bibitem[2003]{Pollack2003} Pollack, L.K., Taylor, G.B., \& Zavala, R.T. 2003, 
                            ApJ, 589, 733

\bibitem[1996]{Reuter1996} Reuter, H.-P., \& Lesch, H. 1996, A\&A, 310, L5

\bibitem[2009]{Reynolds2009} Reynolds, C.S., Nowak, M., Markoff, S., et al. 2009, ApJ, 
                             691, 1159

\bibitem[2008]{Ros2008} Ros, E. \& P\'erez-Torres, M.A. 2008, in ``The role of 
                        VLBI in the Golden Age for Radio Astronomy'', 
                        PoS (IX EVN Symposium) 090

\bibitem[1995]{refdifmap} Shepherd, M.C., Pearson, T.J., \& Taylor, G.B. 1995,
                          BAAS, 27, 2, 903

\bibitem[2007]{Schodel2008} Sch\"odel, R., Krips, M., \& Markoff, S. 
                            2007, A\&A, 463, 551

\bibitem[2011]{Schorr2011} Schorr M\"uller, A., Storchi-Bergmann, T., Riffel, R.A., et al. 2011,
                           MNRAS, 413, 149

\bibitem[2007]{Weiler2007} Weiler, K.W., Williams, C.L., Panagia, N., et al. 
                           2007, ApJ, 671, 1959

\end{thebibliography}
\end{document}